\documentclass[twocolumn,showpacs,preprintnumbers,amsmath,amssymb,floatfix]{revtex4}

\usepackage{graphicx}
\usepackage{dcolumn}
\usepackage{bm}
\begin{document}
\bibliographystyle{plaint}

\preprint{APS/123-QED}

\title{Self-interaction corrected relativistic theory of 
magnetic scattering of x rays: Application to praseodymium}

\author{E. Arola,$^{1,2}$, M. Horne$^1$, P. Strange,$^1$, 
H. Winter,$^3$ Z. Szotek,$^4$ and W.M. Temmerman$^4$}
\affiliation{$^1$School of Chemistry and Physics, Keele University,
Staffordshire ST5 5BG, United Kingdom}
\affiliation{$^2$Optoelectronics Research Centre, 
Tampere University of Technology, P.O. Box 692, 
FIN-33101, Tampere, Finland.}
\affiliation{$^3$INFP, Forschungszentrum Karlsruhe GmbH,
Postfach 3640, D-76021 Karlsruhe, Germany}
\affiliation{$^4$Daresbury Laboratory, Daresbury, Warrington,
WA4 4AD, Cheshire, United Kingdom}

\date{\today}

\begin{abstract}
A first-principles theory of resonant magnetic scattering of x rays is
presented. The scattering amplitudes are calculated using 
a standard time-dependent perturbation theory to second order in the 
electron-photon interaction vertex. In order to calculate the cross 
section reliably an accurate description of the electronic states in 
the material under investigation is required and this is provided by 
the density functional theory (DFT) employing the Local Spin 
Density Approximation combined with the self-interaction corrections 
(SIC-LSD). The magnetic x-ray resonant scattering (MXRS) theory has 
been implemented in the framework of the relativistic spin-polarized 
LMTO-ASA band structure calculation method. The theory is illustrated with
an application to ferromagnetic praseodymium. It is shown that the 
theory quantitatively reproduces the dependence on the spin and 
orbital magnetic moments originally predicted qualitatively (Blume, 
J. Appl. Phys, {\bf 57}, 3615 (1985)) and yields results that can be 
compared directly with experiment. 
\end{abstract}


\maketitle

\section{Introduction}
\label{sec:intro}
Magnetic X-Ray Scattering (MXRS) is a well developed technique for 
probing the magnetic and electronic structures of materials. The 
foundations of the theory of MXRS were laid down by 
Blume. \cite{Blum85} Later on Blume and Gibbs \cite{Blum88} developed 
the theory further to show that the orbital and spin contributions to 
the magnetic moment can be measured separately using MXRS with a 
judicious choice of experimental geometry and polarization of the 
x rays. Hannon \emph{et al.} \cite{Hann88} presented a nonrelativistic 
theory of x-ray resonance exchange scattering and wrote down explicit 
expressions for the electric dipole ($E1$) and quadrupole ($E2$) 
contributions. This work is based on an atomic model of magnetism and 
has been applied successfully to a variety of materials including UAs 
and Gd by Fasolino \emph{et al.}. \cite{Faso93} Rennert \cite{Renn93} 
produced a semi-relativistic theory of MXRS written in terms of 
Green's functions, but no such calculations have been performed. More 
recently, theory based on an atomic model of the electronic structure 
of materials has been written down by Lovesey \cite{Love} and 
co-workers and applied successfully to a variety of materials. Takahashi
{\em et al.} have reported a theory which includes the band structure in 
the calculation of anomalous x-ray scattering. \cite{Taka} A 
first-principles theory of MXRS based on a time-dependent second order 
perturbation theory and density functional 
theory \cite{hohen65,Kohn+Vash} was produced by Arola 
\emph{et al.} \cite{Eero97,Eero01} and applied successfully to several 
transition metal materials. \cite{Eero98} This theory is restricted in 
its range of application because of the limitations imposed by the 
local density approximation to DFT which means that the theory can 
only be applied to simple and transition metal materials. This is 
particularly unfortunate because it is in the rare earth and actinide 
materials that the most exotic magnetism in the periodic table occurs.

In recent years advances in electronic structure calculations beyond 
the local density approximation have broadened the range of materials 
for which numerically accurate electronic structure calculations can 
be performed. In particular the LDA+U method \cite{anis93} and the
self-interaction corrected local spin density approximation to density
functional theory \cite{Fuji95,Dzid93,Bris97,Svan90} have met with considerable 
success in describing materials with localized electrons. The latter 
method reduces the degeneracy of the $f$ states at the Fermi level 
and hence also circumvents all the convergence problems associated 
with the LSD approximation to DFT in electronic structure calculations 
for rare earth materials. Notably, the SIC-LSD has provided a very 
good description of the rare earth metal and rare earth chalcogenide 
crystal structures. \cite{nature} A relativistic version of the SIC 
formalism has been derived \cite{Beid97} that has been shown to yield 
an excellent description of the electronic structure of rare earth 
materials in the few cases to which it has been applied. This method 
is reviewed by Temmerman \emph{et al.}. \cite{Bris97}

The fact that electromagnetic radiation can be scattered from the 
magnetic moments of spin-1/2 particles was first shown by Low, and by 
Gell-Mann and Goldberger half a century ago. \cite{Low54} Later on it 
was Platzman and Tzoar \cite{Plat70} who first proposed the use of 
x-ray scattering techniques to study the magnetization density of 
solids. At that time progress in studying magnetic structures using 
x rays was severely hampered because the cross section for magnetic 
scattering is smaller than the cross section for charge 
scattering \cite{Blum85} by a factor of $(\hbar \omega/mc^2)^2$. It was 
Gibbs \emph{et al.} \cite{Gibb88} who first observed a large resonant 
enhancement of the cross section when the energy of the x ray is tuned 
through an absorption edge. Since that time technological advances 
have produced high resolution, high intensity synchrotron radiation 
sources that have transformed magnetic x-ray resonant scattering 
into a practical tool for investigating magnetic and electronic 
structures of materials. Nowadays the world's leading synchrotron 
facilities have beamlines dedicated to this technique \cite{Coop01} and 
applications of resonant x-ray scattering are burgeoning. Reviews of 
the experimental state-of-the-art MXRS techniques have been written 
by  Stirling \cite{Stir99} and Cooper. \cite{Coop99}

Other approaches to interpreting MXRS spectra exist, particularly the 
successful methods based on group theory and angular momentum algebra 
that result in sum rules as described by Borgatti \cite{Laan04} and by 
Carra \cite{Paolo} and Luo. \cite{Luo} The present work should not be 
regarded as a rival theory to these, but rather as an attempt to 
extend the range of density functional methods to describe magnetic 
scattering of x rays in the same way as is done for photoemission and 
other spectroscopies. \cite{Durh84} As a DFT-based theory our work is, 
of course, based on very different approximations to this earlier work, 
making direct comparison between the two theories problematic. 

We have recently implemented a first-principles theory of MXRS that 
is based on a standard time-dependent perturbation theory where the 
scattering amplitudes are calculated to second order in the 
electron-photon interaction vertex. To describe MXRS from a given 
material it is necessary to have an accurate description of the 
electronic structure of the material in question. This is provided by 
using the SIC within the LSD approximation to the density functional 
theory which is implemented using the relativistic spin-polarized 
LMTO-ASA band structure calculation method. \cite{tblmto} The theory 
of MXRS is equivalent to that of Arola {\it et al.}, \cite{Eero97} but 
has been rewritten in a form that is appropriate for implementation 
in connection with the LMTO-ASA method where there is substantial 
experience of SIC methods. The major step forward reported in this 
paper is the integration of the SIC into the MXRS theory which 
enables us to describe rare earth and actinide materials on an equal 
footing with transition and simple materials. 

In this paper, we give a detailed description of the MXRS theory and 
illustrate it in a calculation for praseodymium. The results are 
analysed and discussed. Finally we show that the present work is 
consistent with the earlier theory and demonstrate how the MXRS 
cross section reflects the properties of these materials.

\section{Theory}

\subsection{The relativistic SIC-LSD formalism}

The SIC-LSD approximation \cite{ZP,Comment} is an {\em ab-initio} electronic 
structure scheme, that is capable of describing localization phenomena in 
solids. \cite{Bris97,Svan90,Dzid93} In this scheme the spurious 
self-interaction of each occupied electron state is subtracted from the 
conventional LSD approximation to the total energy functional, which leads 
to a greatly improved description of static Coulomb correlation effects 
over the LSD approximation. This has been demonstrated in studies of the 
Hubbard model, \cite{Rapid,vogl-hub} in applications to 3$d$ 
monoxides \cite{Dzid93,Svan90} and cuprates, \cite{PRL2,Dzid93} 
$f$-electron systems, \cite{nature,walt3,leon} orbital ordering, \cite{rik} 
metal-insulator transitions \cite{temm4} and solid hydrogen. \cite{SSC}

For many applications it is necessary to account for all relativistic
effects including spin-orbit coupling in an electronic structure
calculation. Relativistic effects become progressively more important as we
proceed to heavier elements. They are also extremely important when we are 
considering properties dependent on orbital moments and their coupling to 
electron spins. 

The relativistic total energy functional in the local spin density 
approximation is
\begin{subequations}
\begin{eqnarray}
E^{\rm LSD}[{\overline n}({\bf r})] &=& E_{\rm kin}[n({\bf r})] + 
U[n({\bf r})] + \int V^{\rm ext}({\bf r})n({\bf r})d^3r \nonumber \\ 
& + & E^{\rm LSD}_{\rm xc}[{\overline n}({\bf r})] - 
\int {\bf B}^{\rm ext}({\bf r})\cdot {\bf m}({\bf r})d^3r,
\label{funct}
\end{eqnarray}
where ${\overline n}({\bf r})=(n^{\uparrow}({\bf r}),n^{\downarrow}({\bf
r}))(\equiv (n({\bf r}),m({\bf r})))$ labels the spin up and spin down 
charge density.
\begin{eqnarray}
E_{\rm kin}[n({\bf r})] & = & \sum_{\Lambda} <\psi_{\Lambda} |{\hat T} | 
\psi_{\Lambda} >, \\
E^{\rm LSD}_{xc}[{\overline n}({\bf r})] & = & \int n({\bf r})\epsilon_{xc}
[{\overline n}({\bf r})]d^3r.
\label{lda}
\end{eqnarray}
\end{subequations}
Here ${\hat T}$ is an operator describing the kinetic energy and rest mass
of the electrons
\begin{eqnarray}
{\hat T} = \frac{c\hbar}{i} {\bm \alpha}\cdot\nabla + mc^2(\beta-I_4), \\
\nonumber
\end{eqnarray}
where ${\bm \alpha}$ and $\beta$ are the usual relativistic
matrices \cite{Stra98}. $U[n({\bf r})]$ represents all two particle
interactions including the Breit interaction. $V^{\rm ext}({\bf r})$ is the
external potential,  ${\bf B}^{\rm ext}({\bf r})$ is an external magnetic field.
The density $n({\bf r})$ and the spin density ${\bf m}({\bf r})$ are given by
\begin{eqnarray}
 n({\bf r}) = \sum_{\Lambda}\psi^{\dag}_{\Lambda}({\bf
r})\psi_{\Lambda}({\bf r}),\\
\label{dens}
\nonumber
\end{eqnarray}
\begin{eqnarray}
 {\bf m}({\bf r}) = -\mu_{\bf B}\sum_{\Lambda}\psi^{\dag}_{\Lambda}({\bf
r})\beta{\bm \sigma}_4\psi_{\Lambda}({\bf r}),\\
\label{mag}
\nonumber
\end{eqnarray}
where ${\bm \sigma}_4$ is the $4 \times 4$ matrix spin operator and $\Lambda$ 
represents the quantum numbers. In Eqs. (\ref{dens}) and (\ref{mag}) below we 
have implied a representation in which spin is a good quantum number and the 
sums are over the occupied states. $\epsilon_{xc}[{\overline n}({\bf r})]$ is 
the exchange correlation energy of a gas of constant density and Eq. (\ref{lda}) 
is the local spin density approximation.
  
If we minimise the functional (\ref{funct}) with respect to changes in the
density and spin density we obtain a Dirac-like equation:
\begin{subequations}
\begin{eqnarray}
(\frac{c\hbar}{i} {\bm \alpha}\cdot\nabla & + & mc^2(\beta-I_4)+V^{\rm eff} 
({\bf r}) \nonumber \\ 
& + &\mu_{\rm B}\beta {\bm \sigma}_4\cdot{\bf B}^{\rm eff}({\bf r})) 
\psi_{\Lambda}({\bf r}) =\epsilon_{\Lambda}\psi_{\Lambda}({\bf r}),
\end{eqnarray}
where 
\begin{eqnarray}
V^{\rm eff}({\bf r}) & = & V^{\rm ext}({\bf r}) + \frac{e^2}{ 4 \pi
\epsilon_0} \int \frac{n({\bf r'})}{\vert {\bf r}-{\bf r'}|}d^3r' \nonumber \\
& + & \frac{\delta E^{\rm LSD}_{xc}[{\overline n}({\bf r})]}{\delta n({\bf r})}, \\
{\bf B}^{\rm eff}({\bf r}) & = & {\bf B}^{\rm ext}({\bf r}) - \frac{\delta  
E^{\rm LSD}_{xc} [{\overline n}({\bf r})]}{\delta {\bf m}({\bf r})}, \\
\nonumber
\end{eqnarray}
\end{subequations}
where ${\overline n}({\bf r})\equiv (n^{\uparrow}({\bf r}), n^{\downarrow}
({\bf r}))(\equiv (n({\bf r}), {\bf m}({\bf r})))$.
The local spin density approximation discussed above provides a very
successful description of a variety of properties of condensed matter, but suffers
from a drawback because it contains self-interactions of the single particle
charges. In an exact theory these spurious self-interactions would precisely
cancel. In the LSD the cancellation is only approximate and in materials
where there are well-localised electrons this can lead to significant
errors. The SIC-LSD approach to this problem is to augment the LSD
functional with an extra term that removes this deficiency. \cite{Beid97}
\begin{subequations}
\begin{equation}
E^{\rm SIC-LSD} = E^{\rm LSD} + E^{\rm SIC},
\end{equation}
where 
\begin{equation}
E^{\rm SIC}[\{{\overline n}_{\gamma}({\bf r})\}]= - \sum_{\gamma} 
(U[n_{\gamma} ({\bf r})] + E_{\rm xc}^{\rm LSD}[{\overline n}_{\gamma}({\bf
r})],
\end{equation}
where ${\overline n}_{\gamma}({\bf r})\equiv (n_{\gamma}^{\uparrow}({\bf r}), 
n_{\gamma}^{\downarrow}({\bf r}))(\equiv (n_{\gamma}({\bf r}), {\bf m}_{\gamma} 
({\bf r})))$ 
and
\begin{equation}
U[n_{\gamma}({\bf r})]=\frac{1}{2}\frac{e^2}{4 \pi \epsilon_0}\int 
\int \frac{n_{\gamma}({\bf r})n_{\gamma}({\bf r}')}{|{\bf r}-{\bf r}'|} 
d^3r'd^3r,
\end{equation}
\begin{equation}
E^{\rm LSD}_{\rm xc}[{\overline n}_{\gamma}({\bf r})] = \int n_{\gamma} 
({\bf r}) \epsilon_{\rm xc}[{\overline n}_{\gamma}({\bf r})]d^3r,
\end{equation}
where $\gamma$ runs over all orbitals that are SI-corrected, and 
\begin{eqnarray}
n_{\gamma}({\bf r}) \equiv \psi^{\dag}_{\gamma}({\bf r})\psi_{\gamma}({\bf r}),\\
\label{dens1}
\nonumber
\end{eqnarray}
\begin{eqnarray}
{\bf m}_{\gamma}({\bf r}) \equiv -\mu_{\bf B}\psi^{\dag}_{\gamma}({\bf r})\beta
{\bm \sigma}_4\psi_{\gamma}({\bf r}),\\
\label{mag1}
\nonumber
\end{eqnarray}
\end{subequations}
For the exchange-correlation term in the SIC energy we need to consider a 
fully spin-polarised electron. The corresponding single particle-like wave 
equation is obtained by taking the functional derivative of $E^{\rm SIC-LSD}$ 
with respect to $\psi^*_{\gamma}({\bf r})$ and we obtain
\begin{subequations}
\begin{eqnarray}
(\frac{c\hbar}{i} {\bm \alpha}\cdot\nabla & + & mc^2(\beta-I_4) + V^{\rm eff}({\bf r})
+\mu_{\rm B}\beta {\bm \sigma}\cdot{\bf B}^{\rm eff}({\bf r}) \nonumber \\
& + & V^{\rm SIC}_{\gamma}({\bf r})) \psi_{\gamma}({\bf r})=
\sum_{\gamma'}\lambda_{\gamma,\gamma'}\psi_{\gamma'} ({\bf r}),
\end{eqnarray}
where the SIC potential is given by
\begin{eqnarray}
& &V^{\rm SIC}_{\gamma}({\bf r})= - \left(\frac{e^2}{4\pi\epsilon_0}\int 
\frac{n_{\gamma}({\bf r'})} {|{\bf r}-{\bf r}'|}d{\bf r}'\right. \nonumber \\ 
& + & \left.\frac{\delta E^{\rm LSD}_{\rm xc}[{\overline n}_{\gamma}({\bf r})]}
{\delta n_{\gamma}({\bf r})}-\mu_{\rm B}\beta {\bm \sigma}_4\cdot
\frac{\delta E^{\rm LSD}_{\rm xc} [{\overline n}_{\gamma}({\bf r})]}
{\delta {\bf m}_{\gamma}({\bf r})}\right).
\end{eqnarray}
\end{subequations}
The task of finding the single particle-like wavefunctions is now
considerably more challenging than for the bare LSD because every state
experiences a different potential. To maintain the orthogonality of the 
$\psi_{\gamma}({\bf r})$ it is necessary to calculate the Lagrange
multiplier matrix, $\lambda_{\gamma \gamma'}$. 

As written in Eqs. (10-13), $E^{\rm SIC-LSD}$ appears to be a functional 
of the set of occupied orbitals rather than of the total spin density 
only, like $E^{\rm LSD}$. By a reformulation it may be shown 
\cite{ZP,Comment} that $E^{\rm SIC-LSD}$ can in fact be regarded as a 
functional of the total spin density only. The associated 
exchange-correlation energy functional 
$E_{\rm xc}^{\rm SIC}[{\overline n}({\bf r})]$ is, however, only 
implicitly defined, \cite{Comment} for which reason  the associated 
Kohn-Sham equations are rather impractical to exploit. For periodic 
solids the SIC-LSD approximation is a genuine extension of the LSD 
approximation in the sense that the self-interaction correction is 
only finite for localized states, which means that if all valence states
considered are Bloch-like single-particle states $E^{\rm SIC-LSD}$ 
coincides with $E^{\rm LSD}$. Therefore, the LSD minimum is also a local 
minimum of $E^{\rm SIC-LSD}$. In some cases another set of single-particle 
states may be found, not necessarily in Bloch form but, of course, equivalent 
to Bloch states, to provide a local minimum for $E^{\rm SIC-LSD}$. For this 
to happen some states must exist which can benefit from the self-interaction 
term without losing too much band formation energy. This usually will be 
the case for rather well localized states like the 3$d$ states in 
transition metal oxides or the 4$f$ states in rare earth compounds. 
Thus, $E^{\rm SIC-LSD}$ is a spin density functional, which may be used 
to describe localized as well as delocalized electron states. 

We have solved the SIC-LSD equations self-consistently for a periodic solid
using the unified Hamiltonian approach described by Temmerman {\em et 
al.}. \cite{Temm3} The equations have been solved on a periodic lattice using
the relativistic LMTO method in the tight-binding representation.

\subsection{The relativistic spin-polarised LMTO method}
 
In Section II C, $u_{\Lambda'}({\bf r})$ will be a general notation 
for the unoccupied intermediate states in the second order time-dependent 
perturbation theory. In the case of a material with translational
periodicity $u_{\Lambda'}({\bf r})$ will be a Bloch state 
\begin{eqnarray}
u_{\Lambda'}({\bf r})=\psi^{j{\bf k}}({\bf r}),
\label{xray.rel.sic.13b}
\end{eqnarray}
for which 
\begin{eqnarray}
\psi^{j{\bf k}}({\bf r}+ {\bf R})= e^{i{\bf k}\cdot{\bf R}}\psi^{j{\bf k}}
({\bf r}),
\label{lmto2}
\end{eqnarray}
where ${\bf k}$ is the wavevector defined to be in the first Brillouin 
zone, $j$ is the band index, and ${\bf R}$ is any Bravais lattice vector. 
In the LMTO method the Bloch wave functions may be expanded in several 
ways. \cite{tblmto} For the calculation of observables it is most convenient 
to make an expansion in terms of the single-site solutions of the radial 
Dirac equation and their energy derivatives. For the relativistic 
spin-polarised case this has been achieved by Ebert \cite{Ebert88,Ebert882} 
and it is this method that we employ. The Bloch state in this representation 
is written as 
\begin{subequations}
\begin{eqnarray}
\psi^{j{\bf k}}({\bf r}) & = & \sum_{t=1}^{N_{\rm type}}\sum_{i=1}^{N_t}
\sum_{\Lambda} \left[\right. A^{j{\bf k}}_{ti\Lambda}\phi_{\nu t \Lambda}
({\bf r}-{\bm \tau}^{(t)}_i) \nonumber \\ 
& + & B^{j{\bf k}}_{ti\Lambda}{\stackrel{.}{\phi}}_{\nu t \Lambda}({\bf
r}-{\bm \tau}^{(t)}_i)\left.\right].
\label{xray.rel.sic.17a}
\end{eqnarray}
Here 
\begin{eqnarray}
\phi_{\nu t \Lambda}({\bf r}^{(t)}_i)=\sum_{\kappa'}\left(\begin{matrix} 
g^{{(t)}m_j}_{ \kappa'\kappa}(\epsilon_{\nu},r^{(t)}_i) \chi^{m_j}_{\kappa'}
({\hat {\bf r}}^{(t)}_i) \cr
if^{{(t)}m_j}_{\kappa'\kappa}(\epsilon_{\nu}, r^{(t)}_i)\chi^{m_j}_{-\kappa'}({\hat 
{\bf r}}^{(t)}_i)\cr \end{matrix}\right),
\label{xray.rel.sic.17b}
\end{eqnarray}
\end{subequations}
where the $g^{(t) m_j}_{\kappa'\kappa}(E,r^{(t)}_i)$ and $f^{(t) m_j}_{\kappa'\kappa}
(E,r^{(t)}_i)$ are solutions of the radial Dirac equation for a spin-polarised
system, and ${\bf r}^{(t)}_i\equiv {\bf r}-{\bm \tau}^{(t)}_i$. Details of the 
solution are given in by Strange {\it et al.}, \cite{Stra84} and 
${\stackrel{.}{\phi}}_{\nu t \Lambda}({\bf r}^{(t)}_i)$ is its energy 
derivative. These satisfy
\begin{eqnarray}
\langle \phi_{\nu t \Lambda}|\phi_{\nu t \Lambda}\rangle =
1 \hskip 15mm \langle \phi_{\nu t \Lambda}|{\stackrel{.}{\phi}}_{\nu 
t \Lambda}\rangle = 0,
\label{xray.rel.sic.18}
\end{eqnarray}
where the subscript $\nu$ corresponds to the energy $\epsilon_{\nu}$ about 
which the muffin-tin orbitals of Eq. (\ref{xray.rel.sic.17b}) are expanded, and the
normalization integrals have been done within the atomic sphere $S^t$. The 
single particle functions $\phi_{\nu t \Lambda}({\bf r})$ and 
${\stackrel{.}{\phi}}_{\nu t \Lambda}({\bf r})$ are evaluated at energy 
$\epsilon_{\nu}$. In this relativistic formulation 
$\Lambda \equiv (\kappa m_j)$ labels the boundary condition for the 
independent single-site solution $\phi_{\nu t \Lambda}({\bf r}-{\bm
\tau}^{(t)}_i)$ of the Dirac equation about the basis atom at ${\bm
\tau}^{(t)}_i$. $N_{\rm type}$ is the number of different types of atom in 
the unit cell. $N_t$ is the number of equivalent atoms of type $t$. The 
coefficients $A^{j{\bf k}}_{ti\Lambda}$ and $B^{j{\bf k}}_{ti\Lambda}$ are 
written in terms of the LMTO structure constants and potential parameters, 
and are completely determined by a self-consistent LMTO calculation of the 
electronic structure. \cite{tblmto} Key observables are then given in terms 
of these quantities. In particular the spin moment is
\begin{subequations}
\begin{eqnarray}
m_S=\sum_j\int_{\epsilon^{j{\bf k}}<\epsilon_{\rm F}}m^{j{\bf k}}_sd^3k,
\end{eqnarray}
where
\begin{eqnarray}
m^{j{\bf k}}_s & = & \sum_{t,i}\sum_{\Lambda}\sum_{\Lambda'}\left( 
 A^{j{\bf k}*}_{ti\Lambda}A^{j{\bf k}}_{ti\Lambda'} 
\langle \phi_{\nu t \Lambda}|\beta \sigma_{4z} | 
\phi_{\nu t \Lambda'}\rangle \right.\nonumber \\
& + &B^{j{\bf k}*}_{ti\Lambda}B^{j{\bf k}}_{ti\Lambda'}\langle 
{\stackrel{.}{\phi}}_{\nu t \Lambda}|\beta \sigma_{4z} |
{\stackrel{.}{\phi}}_{\nu t \Lambda'}\rangle \nonumber \\
& + &A^{j{\bf k}*}_{ti\Lambda}B^{j{\bf k}}_{ti\Lambda'}\langle 
\phi_{\nu t \Lambda}|\beta \sigma_{4z} |
{\stackrel{.}{\phi}}_{\nu t \Lambda'}\rangle \nonumber \\
& + & \left.B^{j{\bf k}*}_{ti\Lambda}A^{j{\bf k}}_{ti\Lambda'}\langle 
{\stackrel{.}{\phi}}_{\nu t \Lambda}|\beta \sigma_{4z} |
\phi_{\nu t \Lambda'}\rangle\right)
\end{eqnarray}
\end{subequations}
with $\epsilon_{\rm F}$ being the Fermi energy and $\epsilon^{j{\bf k}}$ is
the Bloch state eigenenergy. The orbital moment is
\begin{subequations}
\begin{eqnarray}
m_L=\sum_j\int_{\epsilon^{j{\bf k}}<\epsilon_{\rm F}}m^{j{\bf k}}_l
d^3k,
\end{eqnarray}
where
\begin{eqnarray}
m^{j{\bf k}}_l & = & \sum_{t,i}\sum_{\Lambda}\sum_{\Lambda'}\left( \right. 
 A^{j{\bf k}*}_{ti\Lambda}A^{j{\bf k}}_{ti\Lambda'}
\langle \phi_{\nu t \Lambda}|\beta l_z |
\phi_{\nu t \Lambda'}\rangle \nonumber \\ 
& + &B^{j{\bf k}*}_{ti\Lambda}B^{j{\bf k}}_{ti\Lambda'}\langle 
{\stackrel{.}{\phi}}_{\nu t \Lambda}|\beta l_z |
{\stackrel{.}{\phi}}_{\nu t \Lambda'}\rangle \nonumber \\
& + &A^{j{\bf k}*}_{ti\Lambda}B^{j{\bf k}}_{ti\Lambda'}\langle 
\phi_{\nu t \Lambda}|\beta l_z |
{\stackrel{.}{\phi}}_{\nu t \Lambda'}\rangle \nonumber \\
& + & \left.B^{j{\bf k}*}_{ti\Lambda}A^{j{\bf k}}_{ti\Lambda'}\langle 
{\stackrel{.}{\phi}}_{\nu t \Lambda}|\beta l_z |
\phi_{\nu t \Lambda'}\rangle\right)
\end{eqnarray}
\end{subequations}
In all our calculations the {\bf B}-field is along the $z$-axis which
therefore acts as an axis of quantization. 

\subsection{The x-ray scattering cross section}

In this Section we will outline the formal first-principles theory of
magnetic x-ray scattering for materials with translational periodicity. 
The theory is based on the fully relativistic spin-polarised SIC-LMTO 
method in conjunction with 2nd order time-dependent perturbation theory. 
To simplify the presentation a straightforward canonical perturbation 
theory \cite{Stra98} is presented rather than a more sophisticated 
diagrammatic method. \cite{Durh84}  

\subsubsection{Basic theory of x-ray scattering}

The theory of x-ray scattering is based on the second order golden rule for
the transition probability per unit time:
\begin{eqnarray}
w_{if} =\frac{2\pi}{\hbar}\left|\langle f |{\hat H}_{{\rm int}}' | i 
\rangle + \sum_I \frac{\langle f |{\hat H}_{{\rm int}}' | I \rangle
\langle I |{\hat H}_{{\rm int}}' | i \rangle}{E_i-E_I}\right|^2 \nonumber \\ 
\times \delta (E_f-E_i)\\
\nonumber
\label{Gold}
\end{eqnarray}
where $|i\rangle$, $|I\rangle$ and $|f\rangle$ are the initial, intermediate
and final states of the electron-photon system. $E_i$, $E_I$, and $E_f$ are
the corresponding energies. ${\hat H}_{{\rm int}}'$ is the time-independent 
part of the photon-electron interaction operator. The formalism to reduce 
this general expression to single-electron-like form has been published 
previously. \cite{Eero97} Therefore we will not repeat the details here, but 
only the equations that are key to the present implementation. 

In relativistic quantum theory it is the second term in Eq. (\ref{Gold}) 
that is entirely responsible for scattering as it is second order in the 
vector potential. It is convenient to divide this term into four components. 
To see this note that there are just two types of intermediate state 
$|I\rangle$, those containing no photons and those containing two photons. 
We can also divide up the scattering amplitude according to whether or not 
the intermediate states contain excitations from the `negative-energy sea 
of electrons', i.e. the creation of electron-positron pairs. It can be shown
that the x-ray scattering amplitude in the case of elastic scattering can be
written as \cite{Eero97,Eero1} 
\begin{widetext}
\begin{eqnarray}
f{{\bf q}\lambda;{\bf q}'\lambda'}(\omega) & = & 
f^{+{\rm pos}}_{{\bf q}\lambda;{\bf q}'\lambda'}(\omega) + 
f^{-{\rm pos}}_{{\bf q}\lambda;{\bf q}'\lambda'}(\omega) + 
f^{+{\rm neg}}_{{\bf q}\lambda;{\bf q}'\lambda'}(\omega) + 
f^{-{\rm neg}}_{{\bf q}\lambda;{\bf q}'\lambda'}(\omega) \nonumber \\
&=& \sum_{I,\epsilon_\Lambda >0} \frac{\langle f |{\hat H}_{{\rm int}}' | 
I \rangle\langle I |{\hat H}_{{\rm int}}' | i \rangle}{E_i-E_I}
-\sum_{I,\epsilon_\Lambda <0} \frac{\langle f |{\hat H}_{{\rm int}}' | 
I \rangle\langle I |{\hat H}_{{\rm int}}' | i \rangle}{E_i-E_I} \nonumber \\
& = & \sum_{\Lambda \Lambda'}\frac{\int d^3ru^{\dag}_{\Lambda}({\bf
r})X_{{\bf q}'\lambda'}^{\dag}({\bf r})u_{\Lambda'}({\bf r})\int d^3r' 
u^{\dag}_{\Lambda'}({\bf r}')X_{{\bf q}\lambda}({\bf r}')u_{\Lambda}({\bf
r}')}{\epsilon_{\Lambda}-\epsilon_{\Lambda'}+\hbar \omega}\hskip 1cm [1] \nonumber \\
& + & \sum_{\Lambda \Lambda'}\frac{\int d^3ru^{\dag}_{\Lambda}({\bf
r})X_{{\bf q}\lambda}({\bf r})u_{\Lambda'}({\bf r})\int d^3r' 
u^{\dag}_{\Lambda'}({\bf r}')X_{{\bf q}'\lambda'}^{\dag}({\bf r}')u_{\Lambda}({\bf
r}')}{\epsilon_{\Lambda}-\epsilon_{\Lambda'}-\hbar \omega}\hskip 1cm [2]\nonumber \\
& - & \sum_{\Lambda \Lambda'}\frac{\int d^3rv^{\dag}_{{\overline\Lambda}}
({\bf r})X_{{\bf q}'\lambda'}^{\dag}({\bf r})u_{\Lambda'}({\bf r})\int d^3r' 
u^{\dag}_{\Lambda'}({\bf r}')X_{{\bf q}\lambda}({\bf r}')v_{{\overline \Lambda}} 
({\bf r}')}{\epsilon_{\Lambda}-\epsilon_{\Lambda'}+\hbar \omega}\hskip 1cm [3] \nonumber\\
& - & \sum_{\Lambda \Lambda'}\frac{\int d^3rv^{\dag}_{{\overline 
\Lambda}}({\bf r})X_{{\bf q}\lambda}({\bf r})u_{\Lambda'}({\bf r})\int d^3r' 
u^{\dag}_{\Lambda'}({\bf r}')X_{{\bf q}'\lambda'}^{\dag}({\bf r}')v_{{\overline 
\Lambda}}({\bf r}')}{\epsilon_{\Lambda}-\epsilon_{\Lambda'}-\hbar \omega}\hskip 1cm
[4],
\label{pert}
\end{eqnarray}
\end{widetext}
where $u_{{\Lambda}}({\bf r})$ and $v_{{\Lambda}}({\bf r})$ are positive-energy
electron and positron eigenstates of the Dirac Hamiltonian for the crystal
and form a complete orthonormal set of four-component basis functions in the
Dirac space. The quantum state label $\overline{\Lambda}$ can then be
related by symmetry arguments to $\Lambda$. In Eq. (\ref{pert}) term [1] represents 
scattering with no photons and positive energy electrons only in the intermediate 
state, term [2] is when there are two photons and positive energy electrons only 
in the intermediate state, term [3] is for no photons and when negative-energy 
electrons exist in the intermediate state and term [4] is for when two photons and 
negative-energy electrons exist in the intermediate state. We may recall
that within the golden rule based Thomson scattering formalism the negative-energy
related state terms have the wrong sign. Therefore amplitudes $[3]$ and $[4]$ have
been non-rigorously corrected by multiplying them by $-1$. The positive energy 
one-electron states are subject to the constraint that $\epsilon_{\Lambda}\leq 
\epsilon_{\rm F}$ and $\epsilon_{\Lambda'}>\epsilon_{\rm F}$. The relativistic 
photon-electron interaction vertex is
\begin{eqnarray}
X_{{\bf q} \lambda}({\bf r})=-e\left(\frac{\hbar c^2}{2V\epsilon_0\omega}
\right)^{1/2}{\bm \alpha}\cdot {\hat {\epsilon}}^{(\lambda)}({\bf q}) 
e^{i{\bf q}\cdot{\bf r}}
\label{xray.rel.sic.2b}
\end{eqnarray}
where $e=-|e|$, and ${\bf q},\lambda~ ({\bf q}',\lambda')$ represent the 
wavevector and polarisation of the incident (outgoing) photon, and 
${\hat {\bm \epsilon}}^{(\lambda)}({\hat {\bf q}})$ is the polarization vector for 
the x ray propagating in the direction of ${\bf q}$. The ${\bm \alpha}\equiv
(\alpha_x,\alpha_y,\alpha_z)$ are the usual relativistic matrices in the 
standard representation. In Eq.~(\ref{pert}) the last two terms are neglected. The 
justification for this is twofold. Firstly, in the energy range of interest $\hbar 
\omega <<2mc^2$ these two terms have no resonance, and so will only make a 
contribution to the cross section that is slowly varying. This is to be compared 
with the resonant behaviour of the first term. Secondly, in Thomson scattering, 
where the negative energy states play a key role, all the electron states are 
extended. In a crystalline environment the negative energy states are largely 
extended while the states close to the Fermi energy are more 
localised, so one would expect the matrix elements to be smaller. For
further details see Section II C of Ref.\ \onlinecite{Eero97}. 
Henceforth the first term in Eq. (\ref{pert}) will be referred to as 
the resonant term and the second as the non-resonant term. 

In elastic scattering of x rays $u_{\Lambda}({\bf r})$ is an atomic-like core
state localised at a lattice site. Although it is localised it is still an
electron state of the crystal Hamiltonian. It is given by
\begin{eqnarray}
u^{(n)}_{\Lambda_n}({\bf r}_n)=\sum_{\kappa'_n}\left(\begin{matrix} 
g^{(n)m_j}_{\kappa'_n \kappa_n}(r_n) \chi ^{m_j}_{\kappa'_n}({\hat 
{\bf r}}_n) \cr
if^{(n)m_j}_{\kappa'_n\kappa_n}(r_n) \chi^{m_j}_{-\kappa'_n}({\hat {\bf
r}}_n) \cr \end{matrix}\right),
\label{xray.rel.sic.12}
\end{eqnarray}
where $g^{(n)m_j}_{\kappa'_n\kappa_n}(r_n)$ and
$f^{(n)m_j}_{\kappa'_n\kappa_n}(r_n)$ are solutions of the radial 
spin-polarised Dirac equation \cite{Stra84} at the site $n$ and 
$\chi^{m_j}_{\kappa}({\hat {\bf r}})$ are the usual spin-angular functions 
with angular momentum related quantum numbers 
$\Lambda\equiv(\kappa m_j).$ \cite{Stra98,Rose,Eero91} As in Eq.
(\ref{xray.rel.sic.17b}) the 
sum over $\kappa'_n$ runs over $\kappa'_n=\kappa_n$ and $\kappa'_n=-\kappa_n-1$ only.

\subsubsection{\label{sec:amplitudes}Evaluation of the cross section}

The physical observable measured in MXRS experiments is the elastic
differential cross section for scattering.
This is given by (see Sec. II E of Ref.\ \onlinecite{Eero97})
\begin{equation}
\frac{d\sigma}{d\Omega}=
\frac{V^2\omega^2}{(2\pi)^2\hbar^2c^4}
\left| f_{{\bf q}\lambda;{\bf q}^\prime\lambda^\prime}
(\omega)\right|^2,
\label{xray.rel.sic.1c}
\end{equation}
where the symbols have their usual meanings, and we need to calculate
the first two terms of Eq.\ (\ref{pert}), i.e.
$f^{+(\text{pos})}$ and $f^{-(\text{pos})}$.

When implementing Eq.\ (\ref{pert}) for a perfect,
translationally periodic multi-atom per unit cell crystal we use
the following coordinate transformations
\begin{subequations}
\label{xray.rel.sic.19}
\begin{eqnarray}
{\bf r} &\equiv& {\bf R}_{I_i}+{\bf r}_{I_i}=
{\bf R}_i^{(0)}+{\bf R}_I+{\bf r}_{I_i},\\
{\bf r}^\prime &\equiv& {\bf R}_{J_j}+{\bf r}_{J_j}=
{\bf R}_j^{(0)}+{\bf R}_J+{\bf r}_{J_j},
\end{eqnarray}
where ${\bf R}_i^{(0)}$ and ${\bf R}_j^{(0)}$ denote
the $i^{\text{th}}$ and $j^{\text{th}}$ basis atoms,
respectively, in the $0^{\text{th}}$ unit cell, and
${\bf R}_I$ and ${\bf R}_J$ are Bravais lattice vectors.

Furthermore, we use the substitutions
\begin{eqnarray}
\sum_\Lambda &\rightarrow& \sum_N\sum_n\sum_{\Lambda(N_n)},\\
\sum_{\Lambda^\prime} &\rightarrow& \sum_{j{\bf k}},
\end{eqnarray}
and
\begin{eqnarray}
\int_\infty d^3r &\rightarrow& \sum_I\sum_i
\int_{S^{I_i}}d^3r_{I_i},\\
\int_\infty d^3r^\prime &\rightarrow& \sum_J\sum_j
\int_{S^{J_j}}d^3r_{J_j},
\end{eqnarray}
\end{subequations}
where $N$, $I$ and $J$ stand for the label of unit cells,
$n$, $i$, $j$ stand for the label of basis atoms, and
$\Lambda(N_n)$ is the initial core state label for an
atom at site ${\bf R}_{N_n}$. 

Using Eq.\ (\ref{xray.rel.sic.19}) 
and Eq.\ (\ref{xray.rel.sic.13b}) in connection with
term [1] of Eq.\ (\ref{pert}) the resonant part of the
positive-energy scattering amplitude for a perfect crystal
can be written as

\begin{widetext}

\begin{eqnarray}
f_{{\bf q}\lambda;{\bf q}^\prime\lambda^\prime}^{+(\text{pos})}
(\omega) = & & \sum_N\sum_n\sum_{\Lambda(N_n)}\sum_{j{\bf k}}
\frac{1}
{\epsilon_{\Lambda(N_n)}-\epsilon^{j{\bf k}}+\hbar\omega}\nonumber\\
&\times& \sum_I\sum_i\int_{S^{I_i}}d^3r_{I_i}
u_{\Lambda(N_n)}({\bf R}_{I_i}+{\bf r}_{I_i})
X_{{\bf q}^\prime\lambda^\prime}^\dagger({\bf R}_{I_i}+{\bf r}_{I_i})
\psi^{j{\bf k}}({\bf R}_{I_i}+{\bf r}_{I_i})\nonumber\\
&\times& \sum_J\sum_j\int_{S^{J_j}}d^3r_{J_j}^\prime
{\psi^{j{\bf k}}}^\dagger({\bf R}_{J_j}+{\bf r}_{J_j}^\prime)
X_{{\bf q}\lambda}({\bf R}_{J_j}+{\bf r}_{J_j}^\prime)
u_{\Lambda(N_n)}({\bf R}_{J_j}+{\bf r}_{J_j}^\prime),
\label{xray.rel.sic.20}
\end{eqnarray}
\end{widetext}
where the sums are restricted such that
$\epsilon_{\Lambda(N_n)}\le\epsilon_{\text{F}}$ and
$\epsilon^{j{\bf k}} > \epsilon_{\text{F}}$.

We approximate Eq.\ (\ref{xray.rel.sic.20}) in a similar way
as we did earlier in our R-SP-GF-MS method based
MXRS theory (see Section II B of Arola \emph{et al.} \cite{Eero97}).
Because the core states $u_{\Lambda(N_n)}$ participating
to the x-ray scattering (XS) are well-localized around site
${\bf R}_{N_n}$, the dominant contribution to
XS in Eq.\ (\ref{xray.rel.sic.20}) becomes from the
$I_i=N_n$ (i.e. $I=N$, $i=n$) and $J_j=N_n$ (i.e. $J=N$, $j=n$)
terms. From the physical viewpoint, this refers to the situation
where in the anomalous scattering process of x rays a core
electron will be annihilated and created at the same atomic
site (site-diagonal scattering).

Furthermore, we note that in the perfect crystal case the
following properties can be used:
1) ${\bf R}_{N_n}={\bf R}_n^{(0)} + {\bf R}_N$;
2) electronic coordinate ${\bf r}_{N_n}$ can be replaced
by ${\bf r}_n$;
3)  $\psi^{j{\bf k}}({\bf R}_n^{(0)} + {\bf R}_N+{\bf r}_n)=
e^{i{\bf k}\cdot{\bf R}_N}\psi^{j{\bf k}}({\bf R}_n^{(0)} + {\bf r}_n)$,
i.e. Bloch's theorem for intermediate states; and
4) the core state label $\Lambda(N_n)=\Lambda_n$,
i.e. is unit cell independent.
If we also use the explicit form of the photon-electron interaction
vertex of Eq.\ (\ref{xray.rel.sic.2b}) then we end up to the
following expression for the $f^{+(\text{pos})}$ scattering amplitude:
\begin{subequations}
\label{xray.rel.sic.22}
\begin{eqnarray}
& &f_{{\bf q}\lambda;{\bf q}^\prime\lambda^\prime}^{+(\text{pos})}
(\omega) = \sum_N\left\{ 
\sum_{j{\bf k}}\sum_n\sum_{\Lambda_n}
e^{-i({\bf q}^\prime - {\bf q})\cdot{\bf R}_n^{(0)}}\right. \nonumber \\
&\times & \left.
\frac{m_{\Lambda_n}^{(n)+j{\bf k}}({\bf q}^\prime\lambda^\prime)
m_{\Lambda_n}^{(n)+j{\bf k}*}({\bf q}\lambda)}
{\epsilon^{(n)}_{\Lambda_n}-\epsilon^{j{\bf k}}+\hbar\omega}\right\}
e^{-i({\bf q}^\prime - {\bf q})\cdot{\bf R}_N},\nonumber\\
\label{xray.rel.sic.22.1}
\end{eqnarray}
where the resonant matrix elements are defined as
\begin{equation}
m_{\Lambda_n}^{(n)+j{\bf k}}({\bf q}\lambda)\equiv
\int_{S^n}d^3r_n u_{\Lambda_n}^{(n)\dagger}({\bf r}_n)
X_{{\bf q}\lambda}^\dagger ({\bf r}_n)
\psi^{j{\bf k}}({\bf R}_n^{(0)} + {\bf r}_n),
\label{xray.rel.sic.22.2}
\end{equation}
\end{subequations}
where $S^n$ refers to the $n^{\text{th}}$ atomic sphere within
the unit cell.

In Eq.\ (\ref{xray.rel.sic.22})  we notice that
\begin{subequations}
\label{xray.rel.sic.2324}
\begin{equation}
\sum_N e^{-i({\bf q}^\prime - {\bf q})\cdot{\bf R}_N} =
N_{\text{cells}}\sum_{\bf K}\delta_{{\bf q}^\prime - {\bf q},{\bf K}},
\label{xray.rel.sic.23}
\end{equation}
and
\begin{equation}
\sum_{j{\bf k}} \longrightarrow \sum_j\frac{V}{(2\pi)^3}
\int_{1. \text{BZ}} d^3k,
\label{xray.rel.sic.24}
\end{equation}
\end{subequations}
where $N_{\text{cells}}$ is the number of unit cells in the crystal,
${\bf K}$ is a reciprocal lattice vector, and
$\delta_{{\bf q}^\prime - {\bf q},{\bf K}}$ is the Kronecker $\delta$
function.

As the last step, we decompose the general basis atom label $n$ in 
Eq.\ (\ref{xray.rel.sic.22}) into type $t$
($t=1,\ldots,N_{\text{type}}$) and basis atom label $i$
($i=1,\ldots,N_t$), i.e. $n\equiv(ti)$.
Consequently, this introduces the following notational
changes in Eq.\ (\ref{xray.rel.sic.22}):

\begin{eqnarray}
& &\sum_n\rightarrow\sum\limits_{t=1}^{N_{\text{type}}}
\sum\limits_{i=1}^{N_t},
\Lambda_n\rightarrow\Lambda_t,
{\bf R}_n^{(0)}\rightarrow{\bm \tau}_i^{(t)},\nonumber\\
& &\epsilon^{(n)}\rightarrow\epsilon^{(t)},
S^n\rightarrow S^t,
m^{(k)}\rightarrow m^{(ti)},\nonumber\\
& &u^{(k)}\rightarrow u^{(t)},
{\bf r}_k\rightarrow{\bf r}_i^{(t)}.
\label{xray.rel.sic.32}
\end{eqnarray}

Implementing these notations along with
Eq.\ (\ref{xray.rel.sic.2324}) in Eq.\ (\ref{xray.rel.sic.22}),
leads to the final expression for the resonant part of the scattering
amplitude in Bragg diffraction which is
\begin{subequations}
\label{xray.rel.sic.25a33b}
\begin{equation}
f_{{\bf q}\lambda;{\bf q}^\prime\lambda^\prime}^{+(\text{pos})}
(\omega) =
f_{0;{\bf q}\lambda;{\bf q}^\prime\lambda^\prime}^{+(\text{pos})}
(\omega) N_{\text{cells}}\sum_{\bf K}\delta_{{\bf Q}{\bf K}},
\label{xray.rel.sic.25a}
\end{equation}
where the $0^{\text{th}}$ unit cell contribution to the
scattering amplitude is
\begin{eqnarray}
& &f_{0;{\bf q}\lambda;{\bf q}^\prime\lambda^\prime}^{+(\text{pos})}
(\omega)=\sum_j \frac{V}{(2\pi)^3}\int_{{\bf k}\in 1.\text{BZ}} d^3k
\sum\limits_{t=1}^{N_{\text{type}}}
\sum\limits_{i=1}^{N_t}\sum_{\Lambda_t}\nonumber\\
& &\times e^{-i{\bf Q}\cdot{\bm \tau}_i^{(t)}}
\frac{m_{\Lambda_t}^{(ti)+j{\bf k}}({\bf q}^\prime\lambda^\prime)
m_{\Lambda_t}^{(ti)+j{\bf k}*}({\bf q}\lambda)}
{\epsilon_{\Lambda_t}^{(t)} - \epsilon^{j{\bf k}} + \hbar\omega
+ i\Gamma_{\Lambda_t}^{(t)}/2}
\theta(\epsilon^{j{\bf k}} - \epsilon_{\text{F}}),\nonumber\\
& &\label{xray.rel.sic.33b}
\end{eqnarray}
\end{subequations}
where ${\bf Q}\equiv{\bf q}^\prime - {\bf q}$, and the matrix
elements $m_{\Lambda_t}^{(ti)+j{\bf k}}({\bf q}\lambda)$
are given by Eq.\ (\ref{xray.rel.sic.22.2}) with the new
notations of Eq.\ (\ref{xray.rel.sic.32}).
The added
phenomenological parameter $\Gamma_{\Lambda_t}^{(t)}$
represents the natural width of the intermediate states created
by the core hole state $|\Lambda_t\rangle$ at the $t$ type
basis atom.

Similarly, starting from term [2] of Eq.\ (\ref{pert}),
it can be shown that the expression for the nonresonant part of
the scattering amplitude in Bragg diffraction can be written as
\begin{subequations}
\label{xray.rel.sic.29a33c28a}
\begin{equation}
f_{{\bf q}\lambda;{\bf q}^\prime\lambda^\prime}^{-(\text{pos})}
(\omega) =
f_{0;{\bf q}\lambda;{\bf q}^\prime\lambda^\prime}^{-(\text{pos})}
(\omega) N_{\text{cells}}\sum_{\bf K}\delta_{{\bf Q}{\bf K}},
\label{xray.rel.sic.29a}
\end{equation}
where the $0^{\text{th}}$ unit cell contribution to the Bragg
scattering amplitude is
\begin{eqnarray}
& &f_{0;{\bf q}\lambda;{\bf q}^\prime\lambda^\prime}^{-(\text{pos})}
(\omega)=\sum_j \frac{V}{(2\pi)^3}\int_{{\bf k}\in 1.\text{BZ}} d^3k
\sum\limits_{t=1}^{N_{\text{type}}}
\sum\limits_{i=1}^{N_t}\sum_{\Lambda_t}\nonumber\\
& &\times e^{-i{\bf Q}\cdot{\bm\tau}_i^{(t)}}
\frac{m_{\Lambda_t}^{(ti)-j{\bf k}}({\bf q}\lambda)
m_{\Lambda_t}^{(ti)-j{\bf k}*}({\bf q}^\prime\lambda^\prime)}
{\epsilon_{\Lambda_t}^{(t)} - \epsilon^{j{\bf k}} - \hbar\omega}
\theta(\epsilon^{j{\bf k}} - \epsilon_{\text{F}}),\nonumber\\
& &\label{xray.rel.sic.33c}
\end{eqnarray}
where the non-resonant matrix element is defined as
\begin{eqnarray}
m_{\Lambda_t}^{(ti)-j{\bf k}}({\bf q}\lambda) &\equiv&
\int_{S^t}d^3r_i^{(t)}
u_{\Lambda_t}^{(t)\dagger}({\bf r}_i^{(t)})
X_{{\bf q}\lambda}({\bf r}_i^{(t)})\nonumber\\
& &\times\psi^{j{\bf k}}({\bm\tau}_i^{(t)} + {\bf r}_i^{(t)}).
\label{xray.rel.sic.28a}
\end{eqnarray}
\end{subequations}
The total amplitude in Bragg diffraction can then be calculated from
\begin{equation}
f_{{\bf q}\lambda;{\bf q}^\prime\lambda^\prime}^{(\text{pos})}
(\omega) =
\left( f_{0;{\bf q}\lambda;{\bf q}^\prime\lambda^\prime}^{+(\text{pos})}
(\omega) +
f_{0;{\bf q}\lambda;{\bf q}^\prime\lambda^\prime}^{-(\text{pos})}\right)
N_{\text{cells}}\sum_{\bf K}\delta_{{\bf Q}{\bf K}},
\label{xray.rel.sic.33a}
\end{equation}
where $f_0^{+(\text{pos})}$ and $f_0^{-(\text{pos})}$ amplitudes
are given in Eqs.\ (\ref{xray.rel.sic.33b}) and
(\ref{xray.rel.sic.33c}), respectively.

\subsection{\label{sec:matrelems}Matrix elements}

In this section we present the derivation of computable expressions
for the matrix elements
$m_{\Lambda_t}^{(ti)+j{\bf k}}({\bf q}\lambda)$ and
$m_{\Lambda_t}^{(ti)-j{\bf k}}({\bf q}\lambda)$
in the framework of the R-SP-SIC-LMTO electronic structure method.

Using the expansion of Eq.\ (\ref{xray.rel.sic.17a}) for the
SIC-LDA Bloch state $\psi^{j{\bf k}}({\bf r})$, and noticing that
the independent single-site solution $\phi_{\nu t\Lambda}({\bf r})$
of the Dirac equation vanishes outside the atomic sphere at site
$(ti)$, i.e. $\phi_{\nu t\Lambda}({\bf r})=0$ for $r>S^t$
(cf. Ref.\ \onlinecite{Skri84}, pp. 120--1), then the resonant
matrix element $m_{\Lambda_t}^{(ti)+j{\bf k}}({\bf q}\lambda)$
can be written as
\begin{subequations}
\label{xray.rel.sic.35}
\begin{eqnarray}
m_{\Lambda_t}^{(ti)+j{\bf k}}({\bf q}\lambda) &=& \sum_{\Lambda}
\left[ A_{ti\Lambda}^{j{\bf k}}
(u_{\Lambda_t}^{(t)}|X_{{\bf q}\lambda}^\dagger|
\phi_{\nu t\Lambda})\right.\nonumber\\
&+&\left. B_{ti\Lambda}^{j{\bf k}}
(u_{\Lambda_t}^{(t)}|X_{{\bf q}\lambda}^\dagger|
\dot{\phi}_{\nu t\Lambda})\right],
\label{xray.rel.sic.35a}
\end{eqnarray}
where $(f|X_{{\bf q}\lambda}^\dagger|g)$ is defined as
\begin{equation}
(f|X_{{\bf q}\lambda}^\dagger|g)\equiv
\int_{S^t}d^3rf^\dagger({\bf r})
X_{{\bf q}\lambda}^\dagger({\bf r})g({\bf r}),
\label{xray.rel.sic.35b}
\end{equation}
\end{subequations}
where $f\equiv u_{\Lambda_t}^{(t)}$ and
$g\equiv\phi_{\nu t\Lambda}$ or $\dot{\phi}_{\nu t\Lambda}$.

Similarly, it can be shown that the nonresonant matrix
elements can be written as
\begin{eqnarray}
m_{\Lambda_t}^{(ti)-j{\bf k}}({\bf q}\lambda) &=& \sum_{\Lambda}
\left[ A_{ti\Lambda}^{j{\bf k}}
(u_{\Lambda_t}^{(t)}|X_{{\bf q}\lambda}|
\phi_{\nu t\Lambda})\right.\nonumber\\
&+&\left. B_{ti\Lambda}^{j{\bf k}}
(u_{\Lambda_t}^{(t)}|X_{{\bf q}\lambda}|
\dot{\phi}_{\nu t\Lambda})\right],
\label{xray.rel.sic.36a}
\end{eqnarray}
where $(f|X_{{\bf q}\lambda}|g)$ quantities can be calculated
by doing the replacement
$X_{{\bf q}\lambda}^\dagger\rightarrow X_{{\bf q}\lambda}$
in Eq.\ (\ref{xray.rel.sic.35b}).

Finally, we mention few practical points about the implementation
of the matrix elements of  Eqs.\ (\ref{xray.rel.sic.35})
and (\ref{xray.rel.sic.36a}). We will derive below numerically
tractable approximations for these matrix elements due to the
electric dipole ($E1$) or magnetic dipole and electric quadrupole
($M1+E2$) contributions to the photon-electron interaction
vertex $X_{{\bf q}\lambda}({\bf r})$.

\subsubsection{\label{sec:E1}Matrix elements
in electric dipole approximation}

In the electric dipole approximation ($E1$)
[$e^{i{\bf q}\cdot{\bf r}}\approx 1$ in Eq.\ (\ref{xray.rel.sic.2b})],
the resonant matrix element
$m_{\Lambda_t}^{(ti)+j{\bf k}}({\bf q}\lambda)$
of Eq.\ (\ref{xray.rel.sic.35}) can be written as
\begin{subequations}
\label{xray.rel.sic.3938a}
\begin{eqnarray}
ma_{\Lambda_t}^{(ti)+j{\bf k}}({\bf q}\lambda) &=& \sum_{\Lambda}
\left[ A_{ti\Lambda}^{j{\bf k}}
(u_{\Lambda_t}^{(t)}|Xa_{{\bf q}\lambda}^\dagger|
\phi_{\nu t\Lambda})\right.\nonumber\\
&+&\left. B_{ti\Lambda}^{j{\bf k}}
(u_{\Lambda_t}^{(t)}|Xa_{{\bf q}\lambda}^\dagger|
\dot{\phi}_{\nu t\Lambda})\right],
\label{xray.rel.sic.39}
\end{eqnarray}
where
\begin{equation}
Xa_{{\bf q}\lambda}({\bf r})\equiv
-ec\left(\frac{\hbar}{2V\epsilon_0\omega_q}\right)^{1/2}
{\bm\alpha}\cdot\hat{\bm \epsilon}^{(\lambda)}(\hat{\bf q}).
\label{xray.rel.sic.38a}
\end{equation}
\end{subequations}

Using the core state expansion, Eq.\ (\ref{xray.rel.sic.12}), and
the expansions (\ref{xray.rel.sic.17b}) for $\phi_{\nu t\Lambda}
({\bf r})$ and the analogous expansion $\dot{\phi}_{\nu t\Lambda}
({\bf r})$, respectively, it can be shown that
$(u_{\Lambda_t}^{(t)}|Xa_{{\bf q}\lambda}^\dagger|
\phi_{\nu t\Lambda})$
in Eq.\ (\ref{xray.rel.sic.39}) can be written as
\begin{subequations}
\label{xray.rel.sic.40}
\begin{eqnarray}
& &(u_{\Lambda_t}^{(t)}|Xa_{{\bf q}\lambda}^\dagger|
\phi_{\nu t\Lambda}) = -iec\left(
\frac{\hbar}{2V\epsilon_0\omega_q}\right)^{1/2}
\sum_{{\kappa_t}^\prime\kappa^\prime}\nonumber\\
&\times &\left\{ \left[\int_0^{R_{\text{WS}}^{(t)}}dr\,r^2
{g_{{\kappa_t}^\prime\kappa_t}^{(t){m_j}_t}}^*(r)
f_{\kappa^\prime\kappa}^{(t)m_j}(\epsilon_\nu,r)\right]\right. \nonumber \\
&\times& A_{{\kappa_t}^\prime{m_j}_t;-\kappa^\prime m_j}^{(-\lambda)}
(\hat {\bf q})\nonumber\\
&-&\left[\int_0^{R_{\text{WS}}^{(t)}}dr\,r^2
{f_{{\kappa_t}^\prime\kappa_t}^{(t){m_j}_t}}^*(r)
g_{\kappa^\prime\kappa}^{(t)m_j}(\epsilon_\nu,r)\right] \nonumber \\
&\times& \left. A_{-{\kappa_t}^\prime{m_j}_t;\kappa^\prime m_j}
^{(-\lambda)}(\hat {\bf q})\right\},
\label{xray.rel.sic.40a}
\end{eqnarray}
in terms of the radial and angular integrals; $R_{\text{WS}}$
is the Wiegner-Seitz radius and the angular integrals are defined
by \cite{Eero97}
\begin{equation}
A_{\kappa m_j;\kappa^\prime {m_j}^\prime}^{(\lambda)}
(\hat {\bf q})\equiv\int
{\chi_{\kappa}^{m_j}}^{\dagger}(\hat r)
{\bm\sigma}\cdot\hat{\bm \epsilon}^{(\lambda)}({\hat{\bf q}})
\chi_{\kappa^\prime}^{m_j^\prime}(\hat r)d\Omega.
\label{xray.rel.sic.40b}
\end{equation}
\end{subequations}

A numerically tractable expression for
$(u_{\Lambda_t}^{(t)}|Xa_{{\bf q}\lambda}^\dagger|
\dot{\phi}_{\nu t\Lambda})$ in Eq.\ (\ref{xray.rel.sic.39})
can then be written immediately by doing the replacements
$f_{\kappa^\prime\kappa}^{(t)m_j}\rightarrow
\dot{f}_{\kappa^\prime\kappa}^{(t)m_j}$
and
$g_{\kappa^\prime\kappa}^{(t)m_j}\rightarrow
\dot{g}_{\kappa^\prime\kappa}^{(t)m_j}$
on the right side of Eq.\ (\ref{xray.rel.sic.40a}).

Similarly, making a replacement 
$X_{{\bf q}\lambda}\rightarrow Xa_{{\bf q}\lambda}$
in Eq.\ (\ref{xray.rel.sic.36a}), and using the property
$Xa_{{\bf q}\lambda}({\bf r})=
Xa_{{\bf q}-\lambda}^\dagger({\bf r})$ for circularly polarized
light, we can immediately show that the nonresonant matrix
elements can be computed from the resonant ones in the
$E1$ approximation (for further details,
see Sec. II D of Ref.\ \onlinecite{Eero97}) as
\begin{equation}
ma_{\Lambda_t}^{(ti)-j{\bf k}}({\bf q}\lambda)=
ma_{\Lambda_t}^{(ti)+j{\bf k}}({\bf q}-\lambda).
\label{xray.rel.sic.45}
\end{equation}
If the photon propagates along the direction of magnetisation (the $z$-axis)
then the unit polarisation vectors for left (LCP) and right (RCP) circularly
polarised light are ${\hat {\bm \epsilon}}^{(+)}({\hat {\bf z}})=(1,i,0)/\sqrt{2}$ 
and ${\hat {\bm \epsilon}}^{(-)}(\hat{\bf z})=(1,-i,0)/\sqrt{2}$ respectively. 
To obtain the polarisation vectors for propagation directions away from the 
$z$-axis rotation matrices are applied to these vectors. Using the well-known
orthonormality properties of the spherical harmonics the angular integrals
of Eq. (\ref{xray.rel.sic.40b}) can be written
\begin{widetext}
\begin{eqnarray}  
A^{(\lambda )}_{\kappa, m_{j};\kappa'm_j'}(\hat{\bf q}) & = & f_{11}(\theta_q, 
\phi_q,\lambda)C(l\frac{1}{2}j;m_j-\frac{1}{2},\frac{1}{2})C(l'\frac{1}{2}j' 
;m_j'-\frac{1}{2},\frac{1}{2})\delta_{ll'}\delta_{m_j,m_j'} \nonumber \\
 & + & f_{12}(\theta_q, \phi_q,\lambda)C(l\frac{1}{2}j;m_j-\frac{1}{2}, 
\frac{1}{2})C(l'\frac{1}{2}j' ;m_j'+\frac{1}{2},-\frac{1}{2})\delta_{ll'} 
\delta_{m_j,m_j'+1} \nonumber \\
 & + & f_{21}(\theta_q, \phi_q,\lambda)C(l\frac{1}{2}j;m_j+\frac{1}{2}, 
-\frac{1}{2})C(l'\frac{1}{2}j' ;m_j'-\frac{1}{2},\frac{1}{2})\delta_{ll'} 
\delta_{m_j,m_j'-1} \nonumber \\
 & + & f_{22}(\theta_q, \phi_q,\lambda)C(l\frac{1}{2}j;m_j+\frac{1}{2}, 
-\frac{1}{2})C(l'\frac{1}{2}j' ;m_j'+\frac{1}{2},-\frac{1}{2})\delta_{ll'} 
\delta_{m_j,m_j'}. \\
\label{dslr}
\nonumber
\end{eqnarray}
\end{widetext}
The angular factors $f_{ij}(\theta, \phi,\lambda)$ are determined by the
direction of propagation and the photon polarisation. They are discussed in
detail by Arola {\it et al.}. \cite{Eero97} In the case where the direction of
${\bf q}$ is described by a rotation around the $y$-axis of $\theta_q$ followed 
by a rotation about the $z$-axis of $\phi_q$, in the active interpretation they 
are given by
\begin{widetext}
\begin{eqnarray}
\left(\begin{matrix} f_{11} (\theta_q,\phi_q,+) &  f_{12} (\theta_q,\phi_q,+) \cr  
 f_{21} (\theta_q,\phi_q,+) &  f_{22} (\theta_q,\phi_q,+) \cr \end{matrix} \right) = 
\frac{1}{\sqrt{2}}\left(\begin{matrix} -\sin \theta_q  & (\cos \theta_q +1) 
\exp(-i\phi_q) \cr (\cos \theta_q -1)\exp(i\phi_q) & \sin \theta_q
\end{matrix} \right)\\
\left(\begin{matrix} f_{11} (\theta_q,\phi_q,-) &  f_{12} (\theta_q,\phi_q,-) \cr  
 f_{21} (\theta_q,\phi_q,-) &  f_{22} (\theta_q,\phi_q,-) \cr \end{matrix} \right) = 
\frac{1}{\sqrt{2}} \left(\begin{matrix} -\sin \theta_q  & (\cos \theta_q -1)
\exp(-i\phi_q) \cr (\cos \theta_q +1)\exp(i\phi_q) & \sin \theta_q
\end{matrix} \right)\\
\nonumber
\end{eqnarray}
\end{widetext}
for positive and negative helicity x rays, respectively. The angular matrix 
elements of Eq. (\ref{xray.rel.sic.40b}) together with the symmetry of the 
$A^{j{\bf k}}_{ti\Lambda}$ and $B^{j{\bf k}}_{ti\Lambda}$ coefficients 
determines the selection rules in the electric dipole approximation. 

It is important to note that the selection rules, 
derived originally for x-ray scattering in the framework
of the Green's function multiple scattering electronic
structure theory, \cite{Eero97}
can be applied as such only to each term of
Eq.\ (\ref{xray.rel.sic.39}) separately with
angular momentum -like quantum numbers of
the core state ($\Lambda_t$)
and single-site valence orbital ($\Lambda$). \cite{Eero3}
The $E1$ selection rules then become
$l-l_t=\pm 1$ for RCP and LCP radiation in any
propagation direction, while $m_j-{m_j}_t=0,\pm 1$,
depending on the polarization state as well as
on the propagation direction of the photon.
It is also noticeable that the selection rules in
the case of matrix elements
$ma_{\Lambda_t}^{(ti)+j{\bf k}}({\bf q}\lambda)$ are slightly
different from the case of matrix elements
$ma_{\Lambda_t}^{(ti)-j{\bf k}}({\bf q}\lambda)$
with respect to the azimuthal
$m_j$ quantum number, because Eq.\ (\ref{xray.rel.sic.40a})
contains angular matrix elements of the form $A^{(-\lambda)}$,
while the corresponding expression for
$ma_{\Lambda_t}^{(ti)-j{\bf k}}({\bf q}\lambda)$
contains $A^{(\lambda)}$ with an opposite polarization
state index. \cite{Eero97}

Derivation of the selection rules for the matrix elements
$ma_{\Lambda_t}^{(ti)+j{\bf k}}({\bf q}\lambda)$ or
$ma_{\Lambda_t}^{(ti)-j{\bf k}}({\bf q}\lambda)$
would be possible only for ${\bf k}$ points of high
symmetry whose irreducible double point group
representations and the angular momentum
$\Lambda$ ($\kappa m_j$) decomposition for their
symmetrized wavefunctions are known.
However, we apply numerical rather than group theoretical
procedure to determine the selection rule properties
of the abovementioned matrix elements.

\subsubsection{\label{sec:M1+E2}Matrix elements due to
magnetic dipole\protect\\ and electric quadrupole correction}

We derive below an expression for the combined
magnetic dipole and electric quadrupole ($M1+E2$) correction
to the electric dipole
approximation ($E1$) of the matrix elements of
Eqs.\ (\ref{xray.rel.sic.35a}) and
(\ref{xray.rel.sic.36a}). If we now approximate
$e^{i{\bf q}\cdot{\bf r}}\approx 1+i{\bf q}\cdot{\bf r}$ in
Eq.\ (\ref{xray.rel.sic.2b}) for $X_{{\bf q}\lambda}({\bf r})$,
then the term $i{\bf q}\cdot{\bf r}$ is responsible for the
$(M1+E2)$ corrections to the electric dipole approximated
matrix elements
$ma_{\Lambda_t}^{(ti)+j{\bf k}}({\bf q}\lambda)$ and
$ma_{\Lambda_t}^{(ti)-j{\bf k}}({\bf q}\lambda)$,
which we denote as
$mb_{\Lambda_t}^{(ti)+j{\bf k}}({\bf q}\lambda)$ and
$mb_{\Lambda_t}^{(ti)-j{\bf k}}({\bf q}\lambda)$, respectively.

It is then a straightforward matter to show that the matrix element
$mb_{\Lambda_t}^{(ti)+j{\bf k}}({\bf q}\lambda)$, related to the
resonant part of the scattering amplitude, can be written as
\begin{subequations}
\label{xray.rel.sic.4846c}
\begin{eqnarray}
mb_{\Lambda_t}^{(ti)+j{\bf k}}({\bf q}\lambda) &=& \sum_{\Lambda}
\left[ A_{ti\Lambda}^{j{\bf k}}
(u_{\Lambda_t}^{(t)}|Xb_{{\bf q}\lambda}^\dagger|
\phi_{\nu t\Lambda})\right.\nonumber\\
&+&\left. B_{ti\Lambda}^{j{\bf k}}
(u_{\Lambda_t}^{(t)}|Xb_{{\bf q}\lambda}^\dagger|
\dot{\phi}_{\nu t\Lambda})\right],
\label{xray.rel.sic.48}
\end{eqnarray}
where
\begin{equation}
Xb_{{\bf q}\lambda}({\bf r})\equiv
-ec\left(\frac{\hbar}{2V\epsilon_0\omega_q}\right)^{1/2}
{\bm\alpha}\cdot\hat{\bm \epsilon}^{(\lambda)}(\hat {\bf q})
i{\bf q}\cdot{\bf r}.
\label{xray.rel.sic.46c}
\end{equation}
\end{subequations}

Using again the angular momentum expansions of the core state
$u_{\Lambda_t}^{(t)}$, $\phi_{\nu t\Lambda}$,
and $\dot{\phi}_{\nu t\Lambda}$ functions,
as we did in the derivation of the 
$ma_{\Lambda_t}^{(ti)+j{\bf k}}({\bf q}\lambda)$ expression,
we get for the first term of Eq.\ (\ref{xray.rel.sic.48}) as
\begin{subequations}
\label{xray.rel.sic.49}
\begin{eqnarray}
& &(u_{\Lambda_t}^{(t)}|Xb_{{\bf q}\lambda}^\dagger|
\phi_{\nu t\Lambda}) = -ec\left(
\frac{\hbar}{2V\epsilon_0\omega_q}\right)^{1/2}
q\sum_{{\kappa_t}^\prime\kappa^\prime}\nonumber\\
&\times&\left\{  \left[\int_0^{R_{\text{WS}}^{(t)}}dr\,r^3
{g_{{\kappa_t}^\prime\kappa_t}^{(t){m_j}_t}}^*(r)
f_{\kappa^\prime\kappa}^{(t)m_j}(\epsilon_\nu,r)\right] \right. \nonumber \\
&\times& B_{{\kappa_t}^\prime {m_j}_t;-\kappa^\prime m_j}^{(-\lambda)}
(\hat{\bf q})\nonumber\\
&-&\left[\int_0^{R_{\text{WS}}^{(t)}}dr\,r^3
{f_{{\kappa_t}^\prime\kappa_t}^{(t){m_j}_t}}^*(r)
g_{\kappa^\prime\kappa}^{(t)m_j}(\epsilon_\nu,r)\right] \nonumber \\
&\times&\left. B_{-{\kappa_t}^\prime {m_j}_t;\kappa^\prime m_j}
^{(-\lambda)}(\hat {\bf q})\right\},
\label{xray.rel.sic.49a}
\end{eqnarray}
where the angular integrals are defined by \cite{Eero97}
\begin{equation}
B_{\kappa m_j;\kappa^\prime m_j^\prime}^{(\lambda)}(\hat {\bf q}) 
\equiv\int {\chi_{\kappa}^{m_j}}^{\dagger}(\hat {\bf r})
{\bm\sigma}\cdot\hat{\bm \epsilon}^{(\lambda)}({\hat{\bf q}})
\hat{\bf q}\cdot\hat{\bf r}
\chi_{\kappa^\prime}^{m_j^\prime}(\hat {\bf r})d\Omega,
\label{xray.rel.sic.49b}.
\end{equation}
\end{subequations}
where $|\hat {\bf q}|=|\hat {\bf r}|=1$.

A similar expression can be worked out for the second term
of Eq.\ (\ref{xray.rel.sic.48}) by doing the replacements
$f_{\kappa^\prime\kappa}^{(t)m_j}\rightarrow
\dot{f}_{\kappa^\prime\kappa}^{(t)m_j}$
and
$g_{\kappa^\prime\kappa}^{(t)m_j}\rightarrow
\dot{g}_{\kappa^\prime\kappa}^{(t)m_j}$
on the right side of Eq.\ (\ref{xray.rel.sic.49a}).

By Eq.\ (\ref{xray.rel.sic.36a}), the nonresonant matrix
elements $mb_{\Lambda_t}^{(ti)-j{\bf k}}({\bf q}\lambda)$,
due to the ($M1+E2$) correction, can then be worked out
by making the replacement
$Xb_{{\bf q}\lambda}^\dagger\rightarrow Xb_{{\bf q}\lambda}$
in Eq.\ (\ref{xray.rel.sic.48}).
By noticing that $Xb_{{\bf q}\lambda}({\bf r})=
-Xb_{{\bf q}-\lambda}^\dagger({\bf r})$,
we can express the nonresonant ($M1+E2$) matrix elements
in terms of the resonant ones as
\begin{equation}
mb_{\Lambda_t}^{(ti)-j{\bf k}}({\bf q}\lambda)=
-mb_{\Lambda_t}^{(ti)+j{\bf k}}({\bf q}-\lambda).
\label{xray.rel.sic.53}
\end{equation}

The angular matrix elements
$B_{\kappa m_j;\kappa^\prime m_j^\prime}^{(\lambda)}
(\hat q)$ of  Eq.\ (\ref{xray.rel.sic.49b})
can be written as a sum of twelve terms
(see Eq. (26) of Ref.\ \onlinecite{Eero97}).
Consequently, the selection rules of the ($M1+E2$)
contribution to the x-ray scattering are essentially
more complicated than in the $E1$ case.

As guided by the $E1$ case above, we can derive the
($M1+E2$) selection rules for each term of
 Eq.\ (\ref{xray.rel.sic.48}) with
angular momentum -like quantum numbers of the core
state ($\Lambda_t$)
and single-site valence orbital ($\Lambda$).
The resulting selection rules are then
$l-l_t=0,\pm 1,\pm 2$ with the restriction that $s\rightarrow p$
and $p\rightarrow s$ be forbidden transitions, and for the
azimuthal quantum number $m_j-{m_j}_t=0,\pm 1,\pm 2$, depending on
the direction and polarization of the photon. \cite{Eero97}

\section{\label{sec:results}Results}

In this section we discuss a series of calculations to
illustrate the relativistic MXRS theory we have developed within
the SIC-LSD method for ordered magnetic crystals, and to 
demonstrate explicitly what information is contained in the 
x-ray scattering cross section. For this we have chosen to 
examine fcc praseodymium for a detailed analysis of the 
theory. The reasons for this choice are as follows:
(i) Praseodymium contains two localized $f$-electrons. Therefore, it 
is the simplest $f$-electron material for which we can to a large 
extent alter both the spin and orbital contributions to the magnetic 
moment by selectively choosing beforehand for which electrons we apply 
the SIC correction.
(ii) Being ferromagnetic and fcc it has only one atom per primitive cell 
and is therefore computationally efficient to work with. While the fcc
structure is not the observed ground state of Pr, it has been fabricated
with this structure at high temperatures and pressures.
(iii) Using nonrelativistic SIC-LSD we have obtained good agreement
with experiment for the valence and equilibrium lattice constant of 
praseodymium.
(iv) Preliminary calculations indicate that for the rare earth 
$M_{\text{IV}}$ and $M_{\text{V}}$ edges the MXRS spectra are, to 
first order, independent of crystal structure, so the results we 
obtain may be provisionally compared with experiment.

\subsection{Ground state properties}

\begin{figure}
\includegraphics[scale=.45,angle=270]{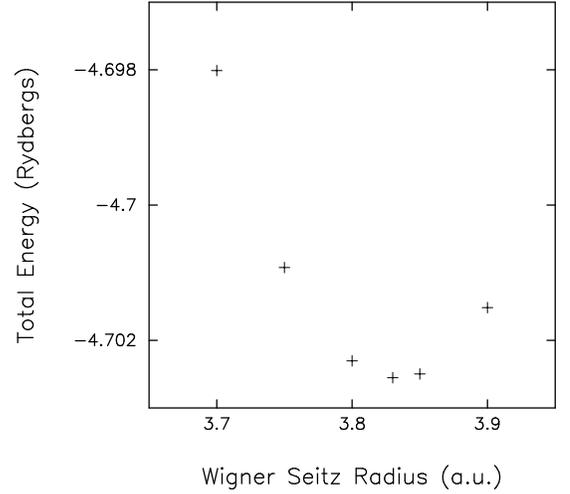}
\caption{The calculated SIC-LSD total energy of fcc praseodymium as a 
function of Wigner-Seitz radius. The electronic configuration 
corresponding to the Hund's rule ground state was used for these 
calculations. The theoretical prediction of the Wigner-Seitz radius is 
3.82 a.u. Experimentally (see text) the value is 3.818 a.u, \cite{tblmto}
or 3.827 a.u. \cite{Bucher} or 3.793 a.u. \cite{Kutz}.}
\label{fig1}
\end{figure} 
We have performed a self-consistent fully relativistic SIC-LSD 
calculation of the electronic structure of praseodymium at a series of 
lattice constants on the fcc structure and found a minimum in the 
total energy as shown in Fig. \ref{fig1}, where the results are presented in 
terms of the Wigner-Seitz radius. There is are variety of different methods
for obtaining the experimental lattice constant. Firstly we can use the 
Wigner-Seitz radius that corresponds to the same volume per atom on the 
fcc lattice as is found in the naturally occuring dhcp crystal
structure \cite{tblmto}. This gives $R_{\rm WS}=3.818$ a.u. Secondly we can 
take the room temperature value which is obtained experimentally from 
flakes of Pr by quenching in an arc furnace. This yields $R_{\rm WS}=3.827$ a.u.
and we can take the value reported by Kutznetsov \cite{Kutz}, $R_{\rm WS}=3.793$ 
a.u. which was measured on samples at 575 K. Clearly our calculated value of
$R_{\rm WS}=3.82$ a.u. is in excellent agreement with these values. Following
earlier work by Myron and Liu \cite{Myron} S\"oderlind performed a 
comprehensive first-principles study of the electronic structure of Pr using
the full potential LMTO method which shows that the fcc phase is stable at
pressures between 60 and 165 kbar \cite{Sod02}. Calculations employing the
SIC within a non-relativistic framework have been performed by
Temmerman {\em et al.} \cite{walt3} and by Svane {\em et al.} \cite{Axel}
\begin{table}
\begin{ruledtabular}
\caption{This table displays the $f$-states selected for self-interaction
correction and the self-consistently calculated spin and orbital 
magnetic moments of those states. The first column simply labels 
different configurations of localized states, the second column gives 
the $m_l$ and $m_s$ quantum numbers of the states from which 
self-interactions have been removed. Columns 3 and 4 are the calculated 
spin and orbital contribution to the total magnetic moment from the 
self-interaction corrected $f$-electrons shown in column 2. Columns 5 
and 6 are the calculated total spin and orbital contribution to the 
magnetic moment from all electrons in fcc Pr. Note that the spin moment 
is fairly constant for all the selected configurations.}
\begin{tabular}{cccccc}
 {N} & {$(m_l,m_s)_1$ $(m_l,m_s)_2$} & {M$_s$({\rm SIC})} &
{$M_l$({\rm SIC})} & {$M_L$(t)} & {$M_S$(t)} \\
 1 & (-3,1/2),\hskip 2mm (-2,1/2) & -4.96 & +1.98 & -4.79 & 2.42 \\
 2 & (-3,1/2),\hskip 2mm(-1,1/2) & -3.95 & +2.00 & -3.91 & 2.43 \\
 3 & (-3,1/2),\hskip 2mm(0,1/2) & -2.99 & +1.99 & -2.97 & 2.43 \\
 4 & (-3,1/2),\hskip 2mm(1,1/2) & -1.96 & +1.96 & -1.97 & 2.48 \\
 5 & (-3,1/2),\hskip 2mm(2,1/2) & -0.98 & +1.99 & -1.03 & 2.48 \\
 6 & (-3,1/2),\hskip 2mm(3,1/2) & -0.005 & +2.00 & -0.05 & 2.52 \\
 7 & (3,1/2),\hskip 2mm(-2,1/2) & 1.00 & +1.99 & 0.89 & 2.48 \\
 8 & (3,1/2),\hskip 2mm(-1,1/2) & 2.02 & +1.97 & 1.84 & 2.50 \\
 9 & (3,1/2),\hskip 2mm(0,1/2) & 3.00 & +1.98 & 2.75 & 2.49 \\
 10 & (3,1/2),\hskip 2mm(1,1/2) & 4.00 & +1.97 & 3.80 & 2.52 \\
 11 & (3,1/2),\hskip 2mm(2,1/2) & 4.99 & +1.99 & 4.69 & 2.54 \\
\end{tabular}
\end{ruledtabular}
\vskip 2mm
\end{table}
\begin{figure}
\includegraphics[scale=.45,angle=270]{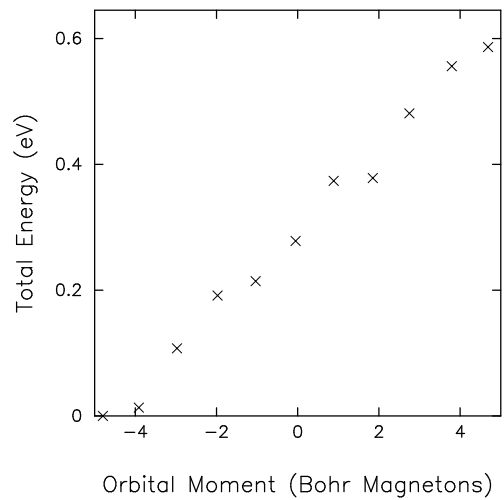}
\caption{The calculated total energy per atom of fcc praseodymium relative
to the ground state energy as a function of orbital moment in Bohr 
magnetons for the states shown in table I. All magnetic moments were 
self-consistently determined and the spin moment was approximately 
constant for all the configurations shown. If an antiparallel arrangement 
of spins was selected the energies were considerably higher.}
\label{fig2}
\end{figure} 

Within the SIC-LSD method we can choose which electron states to 
correct for self-interaction. As the effect of the SIC is to localize 
the states this effectively determines which two of the 14 possible $f$ 
states are occupied in trivalent praseodymium. All non-SI-corrected 
electrons are described using the standard local spin density 
approximation via the unified Hamiltonian describing both localized 
and itinerant electrons. By trying all possible configurations and 
determining which arrangement of $f$ electrons has the lowest total 
energy we can determine the ground state of praseodymium. It should be
pointed out that this interpretation is rather distinct from the 
standard model of the rare earth magnetism where the Hund's rule ground 
state can be thought of as a linear combination of possible 4$f^2$ 
states. In our model the exchange field is automatically included and 
this yields a Zeeman-like splitting of the 4$f^2$ states and gives us 
a unique ground state. In Table I we show a selection of possible 
states occupied by the two electrons with their self-consistently 
evaluated spin and orbital magnetic moments. In Fig. \ref{fig2} we display the 
calculated total energy of these states against orbital moment. Note 
that we have chosen the spin moments parallel for all the states shown. 
For the antiparallel arrangement of electron spins the energy is 
significantly higher. It is clear that there is an approximately linear 
relationship between the total energy and orbital moment. For all the 
points on this figure the orbital as well as spin moments are computed 
self-consistently including the relaxation of the core states. The 
spin moments for all configurations of $f$ electrons are found to be 
approximately the same, always being within 0.06 $\mu_B$ of 2.48 
$\mu_B$ (See Table I) in fair agreement with the result of
S\"oderlind \cite{Sod02}. There is also a small increase in the 
magnitude of the (positive) spin moment as the orbital moment 
increases from its most negative to its most positive values. This 
is due to the increasing effective field felt by the valence 
electrons. There is a slight variation in the spin moment values 
because the small hybridization of the non-SIC corrected $f$ 
electrons with the $5d-6s$ conduction band is dependent on the 
orbital character of the occupied states. Fig. \ref{fig2} is consistent with 
the Hund's rules. The lowest energy state has $f$-spins parallel to 
each other in agreement with Hund's first rule. The total spin moment 
is 2.42$~\mu_{\rm B}$ of which the two localized $f$ electrons contribute 
1.98$~\mu_{\rm B}$, and the remainder comes from spin-polarization in the 
valence bands. The $z$ component of the orbital magnetic moment is 
-4.79$~\mu_{\rm B}$ which is composed of -4.97$~\mu_{\rm B}$ from the 
localized $f$-electrons and 0.18$~\mu_{\rm B}$ from the valence electrons. 
Note that the valence contribution to the orbital moment is parallel to 
the spin moment and antiparallel to the localised orbital moment.These 
numbers are fully consistent with Hund's second rule. Furthermore the 
$f$-shell is less than half full and the spin and orbital moments are 
found to be antiparallel in the lowest energy state, consistent with 
Hund's third rule. The fact that we can reproduce the expected lattice 
constant and $f$-electron configuration suggests very strongly that 
the electronic structure calculated using the relativistic SIC-LSD 
method describes the ground state properties of fcc praseodymium well. 
A detailed discussion of the electronic structure of the rare earth 
metals, calculated using the relativistic SIC-LSD method, will be 
published elsewhere \cite{Paul1}
\begin{figure*}
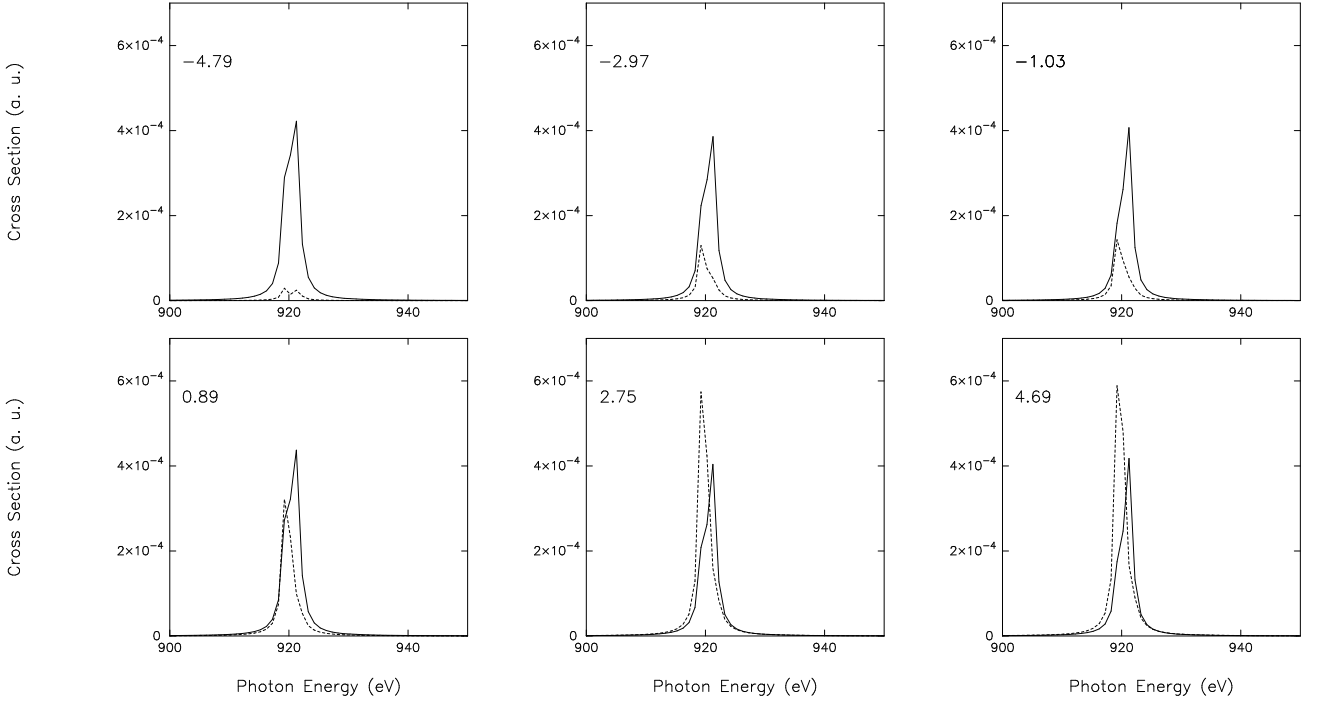

\begin{tabular}{l c c c c c c c c c c c r}
\includegraphics[scale=.33,angle=270]{CSM4M5.ps} & & & & & &
\includegraphics[scale=.33,angle=270]{CSM4M3.ps} & & & & & &
\includegraphics[scale=.33,angle=270]{CSM4M1.ps} \\
\includegraphics[scale=.33,angle=270]{CSM4P1.ps} & & & & & &
\includegraphics[scale=.33,angle=270]{CSM4P3.ps} & & & & & &
\includegraphics[scale=.33,angle=270]{CSM4P5.ps} \\
\end{tabular}
\caption{The scattering cross section at the $M_{\text{IV}}$ edge for 
praseodymium for electron configurations $1,3,5,7,9,$ and $11$ from 
Table I. Each figure is for a different pair of localized 
$f$-electrons. The calculated total orbital moment in Bohr magnetons is 
shown in the top left of each figure. The full curve is the cross 
section for x rays with positive helicity and the dashed curve is that 
for negative helicity x rays. A general trend of increasing magnitude 
of the cross section for negative helicity incident photons as the 
orbital moment increases from negative to positive is clearly observable 
in these curves. The positive helicity curve remains approximately constant
with increasing orbital moment.} 
\label{fig3}
\end{figure*} 

\subsection{X-Ray scattering cross sections}

\begin{figure*}
\begin{tabular}{l c c c c c c c c c c c r}
\includegraphics[scale=.33,angle=270]{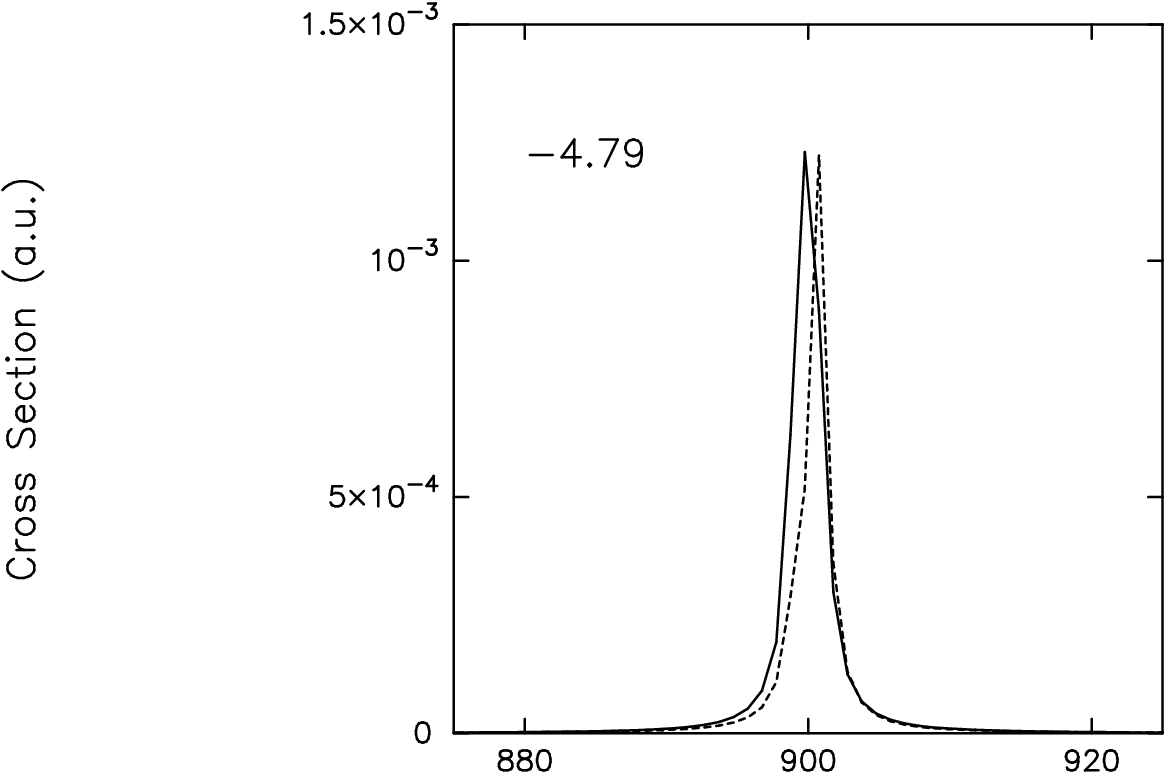} & & & & & &
\includegraphics[scale=.33,angle=270]{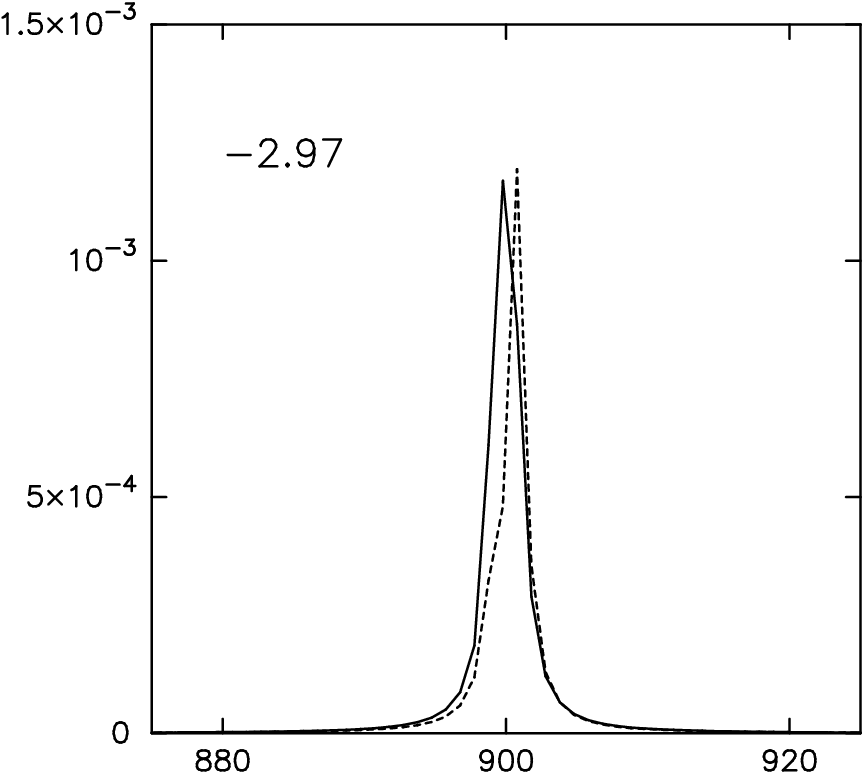} & & & & & &
\includegraphics[scale=.33,angle=270]{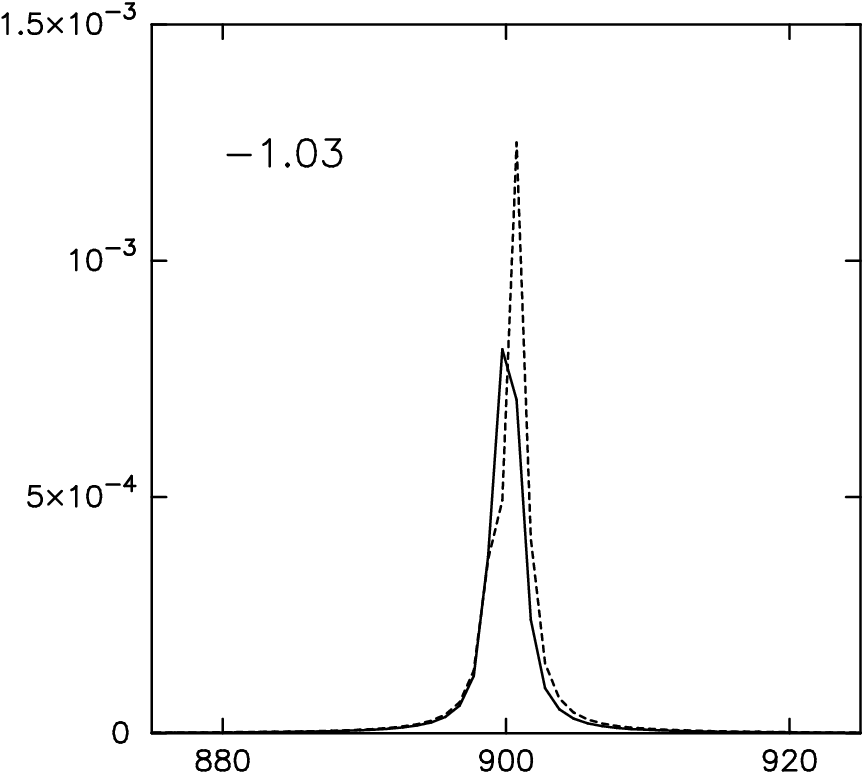} \\
\includegraphics[scale=.33,angle=270]{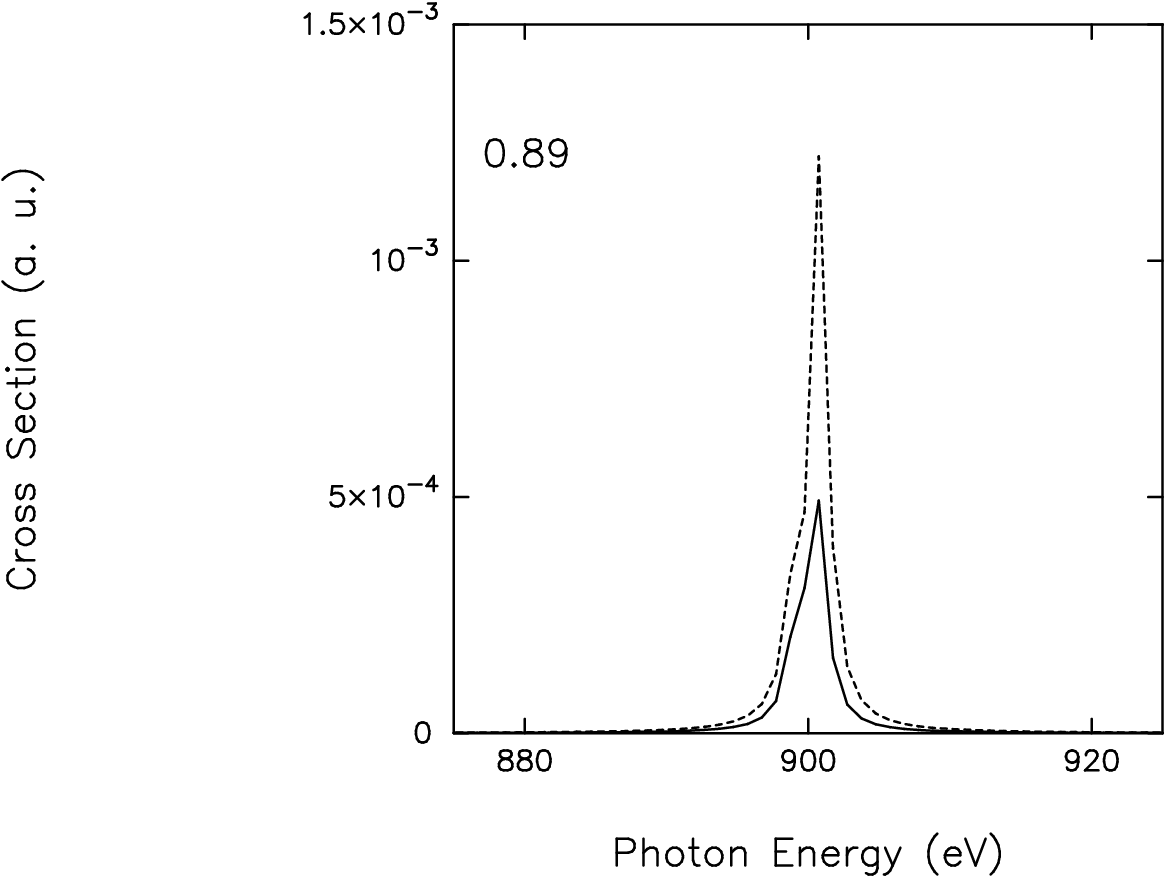} & & & & & &
\includegraphics[scale=.33,angle=270]{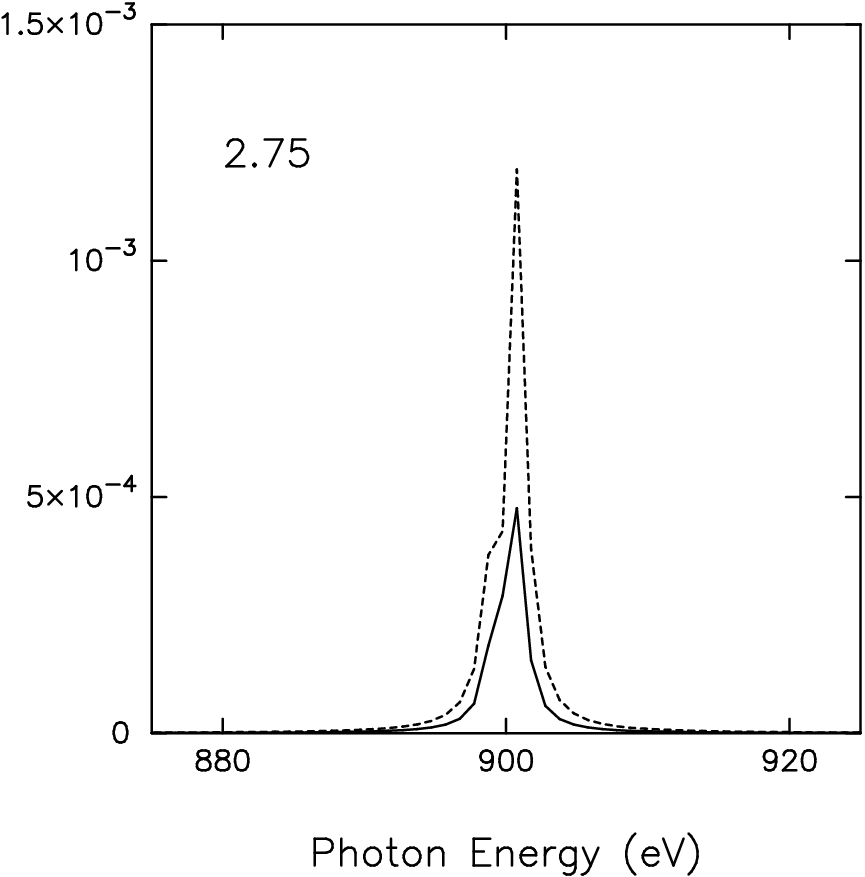} & & & & & &
\includegraphics[scale=.33,angle=270]{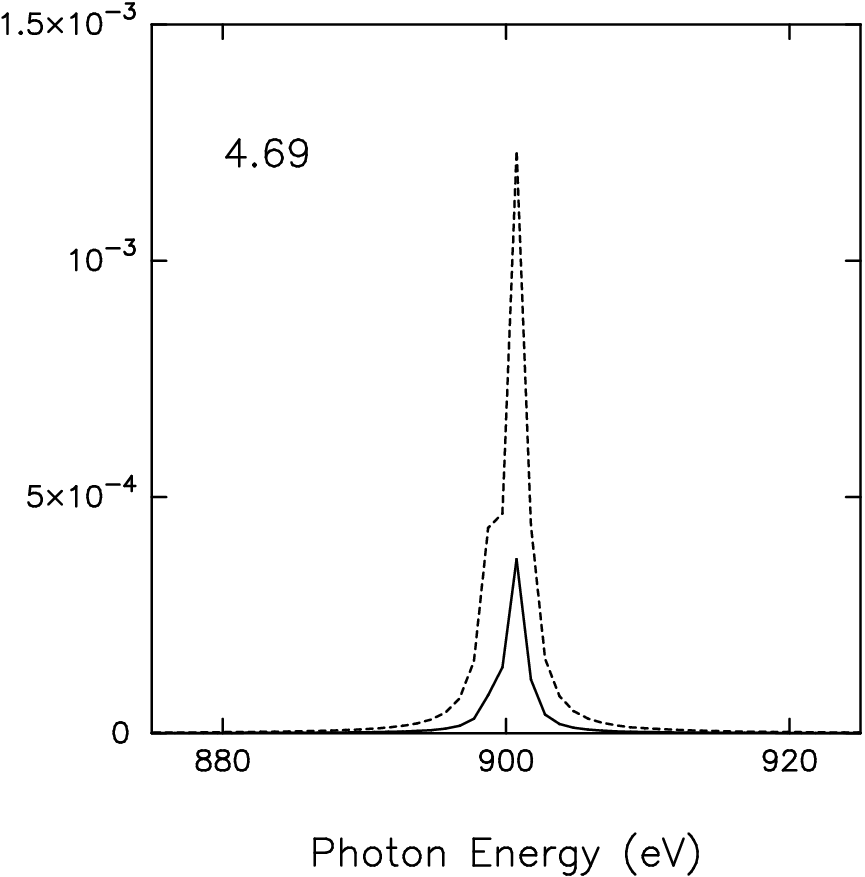} \\
\end{tabular}
\caption{The scattering cross section at the $M_{\text{V}}$ edge for 
praseodymium for electron configurations $1,3,5,7,9$ and $11$ from 
Table I. Each figure is for a different pair of localized 
$f$-electrons. The calculated total orbital moment in Bohr magnetons 
is shown in the top left of each figure. The full curve is the cross 
section for x rays with positive helicity and the dashed curve is that 
for negative helicity x rays. A general trend of decreasing magnitude of 
the cross section for positive helicity incident photons with increasing 
orbital moment is clearly observable in these figures. The negative 
helicity curve remains fairly constant in magnitude with increasing 
orbital moment although the feature on the low energy side of the peak 
does become more pronounced.}
\label{fig4}
\end{figure*}

We have performed calculations of the x-ray scattering cross section 
at the $M_{\text{IV}}$ and $M_{\text{V}}$ absorption edges of Pr for 
all the $f$-electronic configurations shown in Table I. These were 
evaluated with an arbitrary value of $\Gamma$=1~eV which is smaller 
than one would expect experimentally, but as the purpose of this 
section is to investigate the capability of the theory only rather 
than to make a strict comparison with experiment, it does not pose a 
problem. The effect of increasing $\Gamma$ is simply to broaden and 
smooth out the calculated curve. A selection of the results is shown 
in Figs. \ref{fig3} and \ref{fig4}. In each figure the cross section 
for left-handedly circularly polarized (LCP) photons, i.e. with 
positive helicity, and right-handedly circularly polarized (RCP) 
photons, i.e. with negative helicity, are shown. The geometrical setup 
of these calculations assumes that the propagation directions of the 
incident and outgoing photons are parallel to the exchange field, i.e. 
in our case parallel to the spin magnetic moment. Figures \ref{fig3} 
and \ref{fig4} show the cross section at the $M_{\text{IV}}$ and 
$M_{\text{V}}$ edges respectively as the SIC configuration is changed 
systematically such that the $z$ component of the orbital moment varies 
from negative to positive values while at the same time the calculation 
shows that the spin moments remain nearly constant in magnitude and 
parallel to the exchange field. 
As the orbital moment increases we see that the cross section at the 
$M_{\text{IV}}$ edge changes only slightly for LCP x rays while for 
RCP x rays it changes dramatically. At the most negative orbital 
moment the $M_{\text{IV}}$ RCP cross section is very small, being 
completely overshadowed by the LCP peak. At the other end of the 
scale where the orbital moment is most positive the cross section for 
$M_{\text{IV}}$ RCP x rays is considerably larger than that for LCP 
x rays. It should also be noted that the cross section peak for RCP 
x rays is $1-2$ eV lower in energy than the peak for LCP x rays.

When the resonant scattering ($\hbar \omega \approx \epsilon_{\rm F}
-\epsilon_{\Lambda_t}$ ) is close to the $M_{\text{V}}$ edge, it is the 
RCP cross section that remains approximately constant with changing 
orbital moment, although a significant shoulder does appear on the 
low energy side of the curve as the orbital moment increases. The 
LCP peak decreases dramatically with increasing orbital moment. At 
the $M_{\text{V}}$ edge, peaks from RCP and LCP x-ray scattering are 
again separated by $1-2$ eV, but the ordering of the peaks is 
reversed from the case of the $M_{\text{IV}}$ edge scattering.

Figures \ref{fig3} and \ref{fig4} indicate that the $M_{\text{IV}}$ 
and $M_{\text{V}}$ cross sections are directly related to the orbital 
moment of the constituent atoms, although they do not indicate the direct 
proportionality between magnetic moment and scattering cross section 
suggested by Blume \cite{Blum85}. For example, the $M_{\text{V}}$ 
edge cross section for LCP  photons hardly varies in the upper two 
pictures in Fig. \ref{fig4} despite a change of nearly 2$\mu_B$ in the 
orbital moment.
\begin{figure*}
\includegraphics[scale=.33,angle=270]{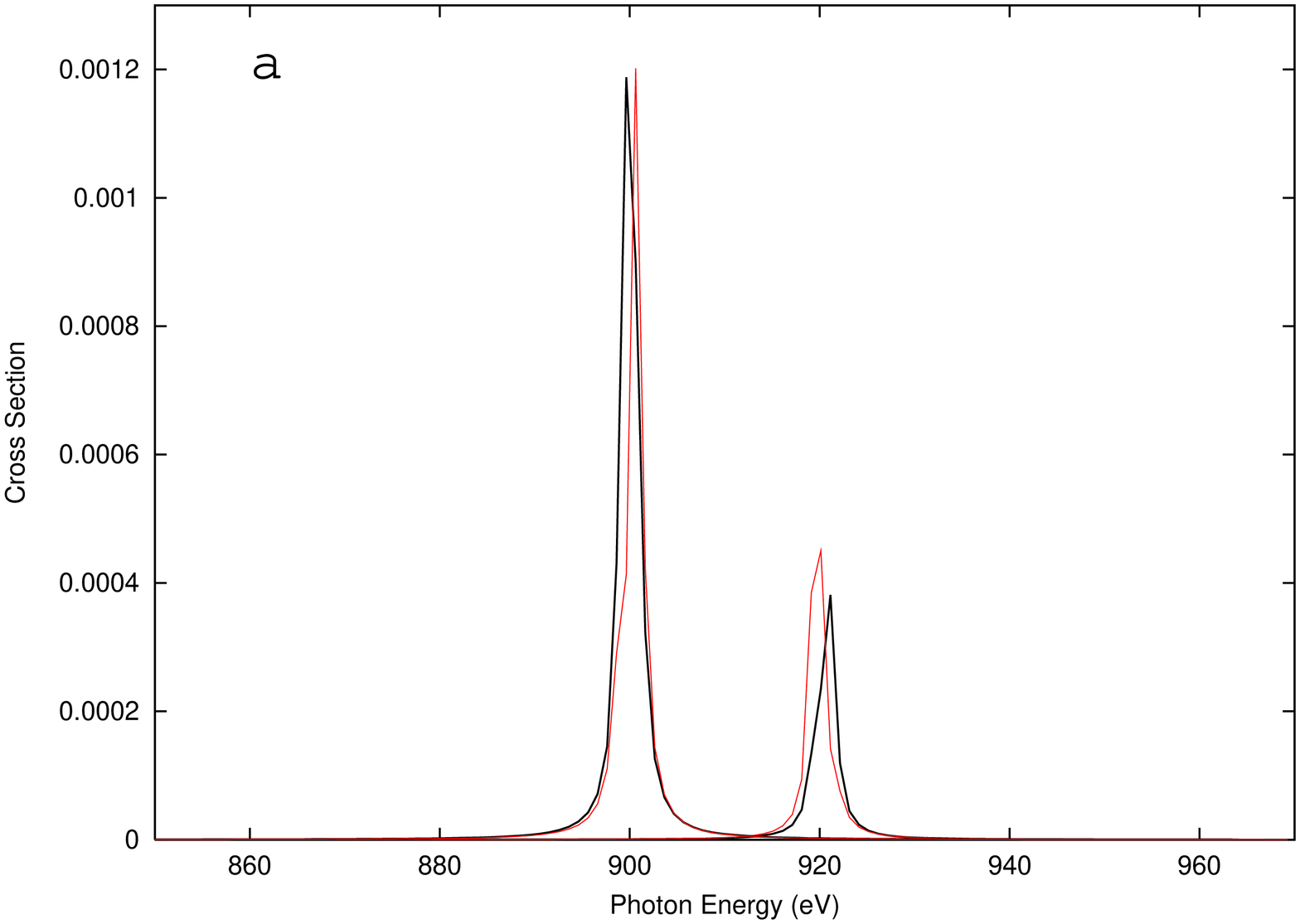}
\includegraphics[scale=.33,angle=270]{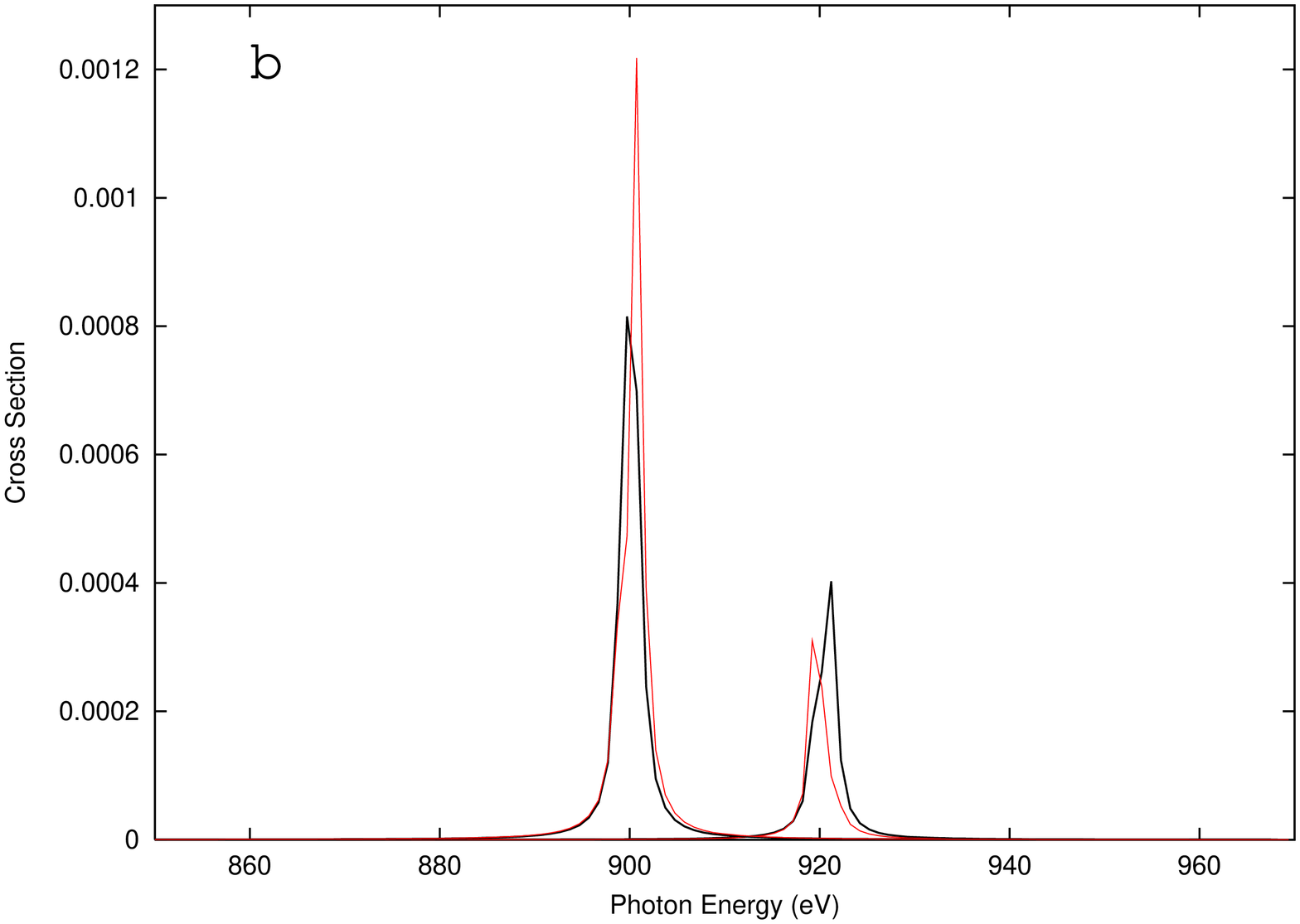}
\includegraphics[scale=.33,angle=270]{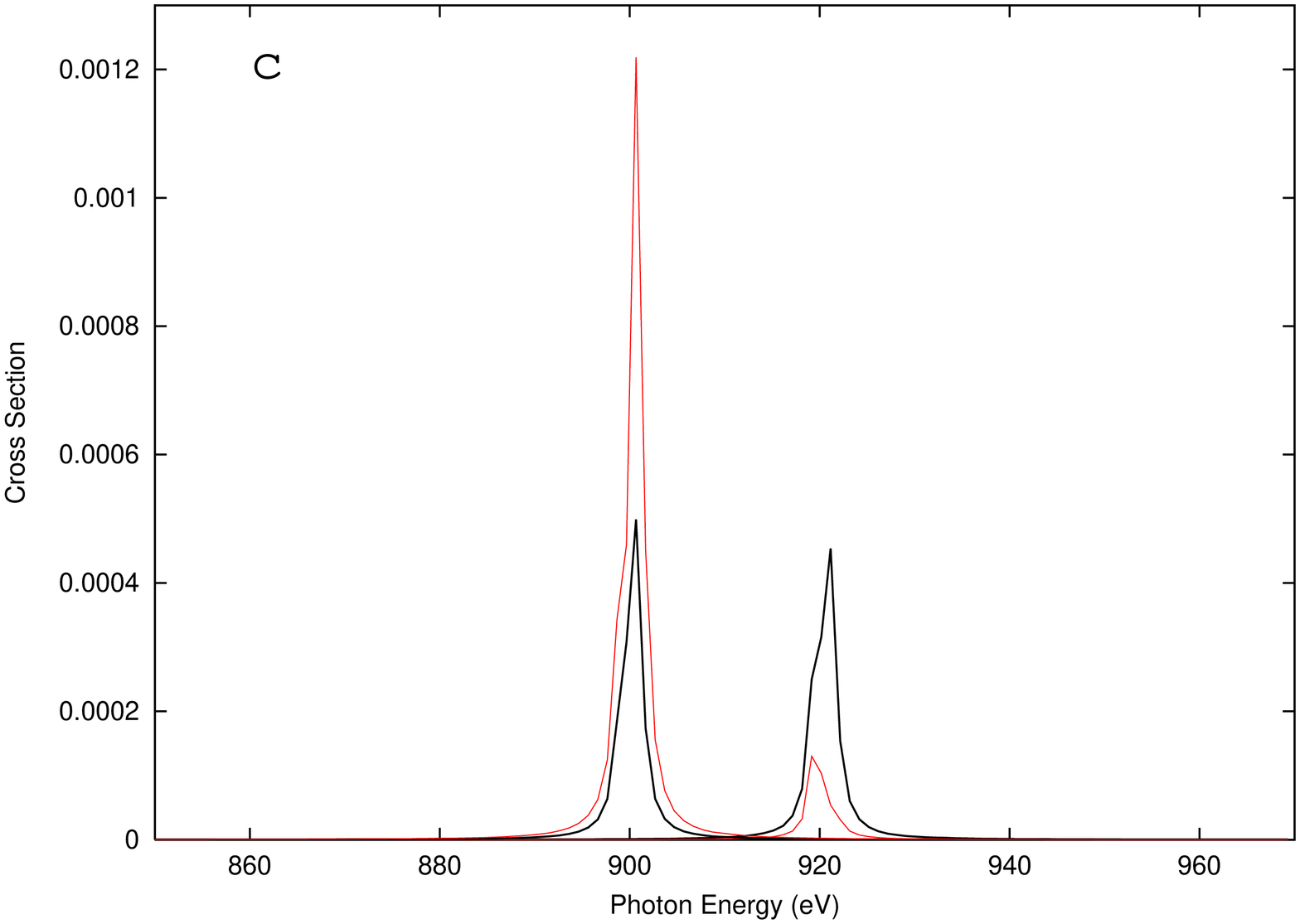}
\caption{The scattering cross section at the $M_{\text{IV}}$ and 
$M_{\text{V}}$ edges for praseodymium. Each figure is for a different 
SIC configuration but which produces roughly the same orbital and 
spin moment. The thick line is for negative helicity incident x rays 
and the thin line is for positive helicity incident x rays:  (a) 
Localising the $m_l=-1$ and $m_l=+1$ electrons with spin up, yields 
an orbital moment of $-0.07~\mu_{\rm B}$ and a spin moment of 
$2.46~\mu_{\rm B}$; 
(b) Localising the $m_l=-2$ and $m_l=+2$ electrons with spin up, 
yields an orbital moment of $-0.12~\mu_{\rm B}$ and a spin moment of 
$2.47~\mu_{\rm B}$; (c) Localising the $m_l=-3$ and $m_l=+3$ electrons with 
spin up, yields an orbital moment of $-0.05~\mu_{\rm B}$ and a spin moment 
of $2.52~\mu_{\rm B}$.}
\label{fig5}
\end{figure*}
To clarify this point further, we show in Fig. \ref{fig5} the cross section 
at the $M_{\text{IV}}$ and $M_{\text{V}}$ edges for SIC configurations 
that produce an orbital moment close to zero with the spins of the 
two occupied $f$ states parallel. While neither the spin nor the 
orbital moment change significantly, the cross section certainly 
does. At the $M_{\text{V}}$ edge the negative helicity curve is 
approximately constant while the positive helicity curve alters 
dramatically. On the other hand, at the  $M_{\text{IV}}$ edge it is
the positive helicity curve that is approximately constant while 
the negative one shows significant variation. This figure implies 
that the resonant x-ray scattering does not measure the total 
orbital moment, but is a measure of the orbital angular momentum of 
the individual one-electron states. 

The important message of Figs. \ref{fig3}--\ref{fig5} is that the scattering 
cross section is not directly proportional to the total orbital moment of 
the material. However, both the spin and orbital moment have a strong 
influence on the size of the cross section peaks.

\section{\label{sec:discussion}Discussion}

The standard theory of x-ray magnetic scattering is based on the work 
of Blume \cite{Blum85}. He derived an equation for the nonresonant
x-ray scattering cross section using a nonrelativistic approach
with relativistic correction to order $1/c^2$. The resulting 
expression for the cross section, using his notation, is
\begin{widetext}
\label{62}
\begin{eqnarray}
\left( \frac{d^2\sigma}{d\Omega^\prime dE^\prime}\right)_{a\to b}
= & &\left( \frac{e^2}{mc^2}\right)^2
\delta(E_a-E_b+\hbar\omega_k-\hbar\omega_{k^\prime}) \nonumber\\
& &\times \left| \langle b|\sum_j e^{i{\bf K}\cdot {\bf r}_j}|a\rangle
~{\bm\epsilon}^\prime\cdot{\bm\epsilon}
-i\frac{\hbar\omega}{mc^2}~
\langle b|\sum_j e^{i{\bf K}\cdot{\bf r}_j}
\left(i\frac{{\bf K}\times{\bf P}_j} {\hbar k^2}\cdot {\bf A}
+ \frac{{\bf s}_j\cdot {\bf B}}{\hbar}\right)|a\rangle\right|^2,
\label{62.a}
\end{eqnarray}
\end{widetext}
where $E_a$ and $E_b$ are the energies of the initial and final
many-electron states, $|a\rangle$ and $|b\rangle$, respectively. 
${\bf K}\equiv{\bf k}-{\bf k^\prime}$, where ${\bf k}$ and 
${\bf k^\prime}$ are the wave vectors of the incoming and 
scattered photons, respectively, and ${\bf r}_j$, ${\bf P}_j$, 
and ${\bf s}_j$ are electron (rigorously density functional state) 
coordinate, momentum, and spin operators.
\begin{equation}
{\bf A} \equiv {\bm\epsilon}^{\prime} \times {\bm\epsilon}
\label{62.b}
\end{equation}
and
\begin{eqnarray}
{\bf B}\equiv {\bm\epsilon}^{\prime} \times {\bm\epsilon}
& - &({\bf{\hat k^\prime}}
\times {\bm\epsilon}^{\prime} ) \times ({\bf{\hat k}} \times
{\bm\epsilon})  \nonumber \\
& - &({\bf{\hat k}} \times {\bm\epsilon})
({\bm\epsilon}^{\prime}\cdot {\bf{\hat k}})+({\bf{\hat k^\prime}}\times
{\bm\epsilon}^{\prime})({\bm\epsilon}\cdot {\bf{\hat k^\prime}})
\label{62.c}
\end{eqnarray}
depend only on the direction and polarization of the incident and 
emitted photons. The first term in  Eq.\ (\ref{62.a}) is the Thomson 
term, responsible for the charge scattering. The term containing 
${\bf A}$ depends on the orbital momentum and the term containing 
${\bf B}$ depends on the electron spin. This expression clearly shows 
that there are three distinct contributions to the magnetic 
scattering cross section, one from the orbital moment, the second 
from the spin moment, and the third from the interference term 
between the spin- and orbital moment. We also note the obvious point 
that if the orbital and spin moments of the individual electrons sum 
to zero, then the magnetic scattering vanishes. Most interestingly,  
Eq.\ (\ref{62.a}) implies that, with a suitable choice of the photon 
energy, geometry and photon polarization it is possible to separate 
contributions to the cross section from the orbital and spin moments.
However, this expression is not directly applicable in our resonant 
magnetic scattering studies because its derivation involves an 
approximation which is strictly not valid close to the resonance, 
while our approach is only valid around resonance because we ignore 
the negative energy contribution to the scattering amplitude, i.e. 
the terms that involve creation of virtual electron-positron pairs 
in the intermediate states in the second order perturbation theory 
(see Sec. II C of Ref.\ \onlinecite{Eero97}). Another difference from 
Blume's theory is the fact that our work is based on fully 
relativistic quantum mechanics, while  Eq.\ (\ref{62.a}) exploits the 
semi-relativistic approximation. This difference makes direct 
comparison of the two theories difficult. This has been discussed by 
Strange \cite{Stra98} who has rederived  Eq.\ (\ref{62.a}) as the 
nonrelativistic limit of a fully relativistic theory of x-ray 
scattering. For these reasons and the fact that there is no 
one-to-one correspondence between the terms in our expression for the 
scattering amplitude and Blume's expression, there is no 
straightforward way to compare the two theories. It is often stated 
that Blume's expression Eq.\ (\ref{62.a}) shows that the cross 
section for magnetic scattering will yield the orbital and spin 
moment of a material separately. Although this will usually be the 
case it is not rigorously true. Eq.\ (\ref{62.a}) cannot be applied 
immediately because the initial and final states $|a>$ and $|b>$ are 
general many-body states that have not been defined in detail. For 
implementation purposes they must be described as many-electron 
states that will contain the index $j$ which is being summed over in 
Eq.\ (\ref{62.a}). In a magnetic material the radial part of the 
basis functions of the single particle wavefunctions, as well as the 
angular part, depend on $m_l$. So, we would expect the total 
scattering amplitude to have a contribution from the orbital angular 
momentum associated with each single-particle state, but this is not 
the same as being proportional to the total orbital angular momentum. 
For example a two-particle state composed of two single-particle 
states with $m_l=\pm 1$ has the same $z$-component of orbital angular 
momentum as a two-particle state composed of two single-particle 
states with $m_l=\pm 3$, but Eq.\ (\ref{62.a}) does not suggest 
that they will have the same scattering amplitude. Nonetheless, 
Blume's expression implies that a strong dependence of the cross 
section on the components of the magnetic moment is likely and 
indeed, this is exactly what we have found, an approximate, but by 
no means rigorous proportionality between orbital moment and 
magnitude of the cross section which is dependent on the 
polarization of the x ray. Furthermore, Figure \ref{fig5} demonstrates 
explicitly the dependence of the cross section on the magnitude of 
$m_l$ of the occupied individual electron states.

The question that now arises is how our computed x-ray scattering
results can be interpreted in terms of the detailed electronic
structure of praseodymium. In order to understand this we analyze 
the electronic structure of fcc Pr for the cases where the orbital 
moment is equal to -4.79 $\mu_B$, -0.05 $\mu_B$, and 4.69 $\mu_B$ 
in detail. We expect the scattering cross section to reflect
the Pr $f$-electron density of states. Although the shape of the
cross section is partially determined by the density of states (DOS) 
the total DOS changes very little when pairs of electrons with 
differing orbital moments are localized. Therefore, a simple 
interpretation of the changes in the cross section with the orbital 
moment in terms of the total DOS cannot be made.
\begin{figure*}
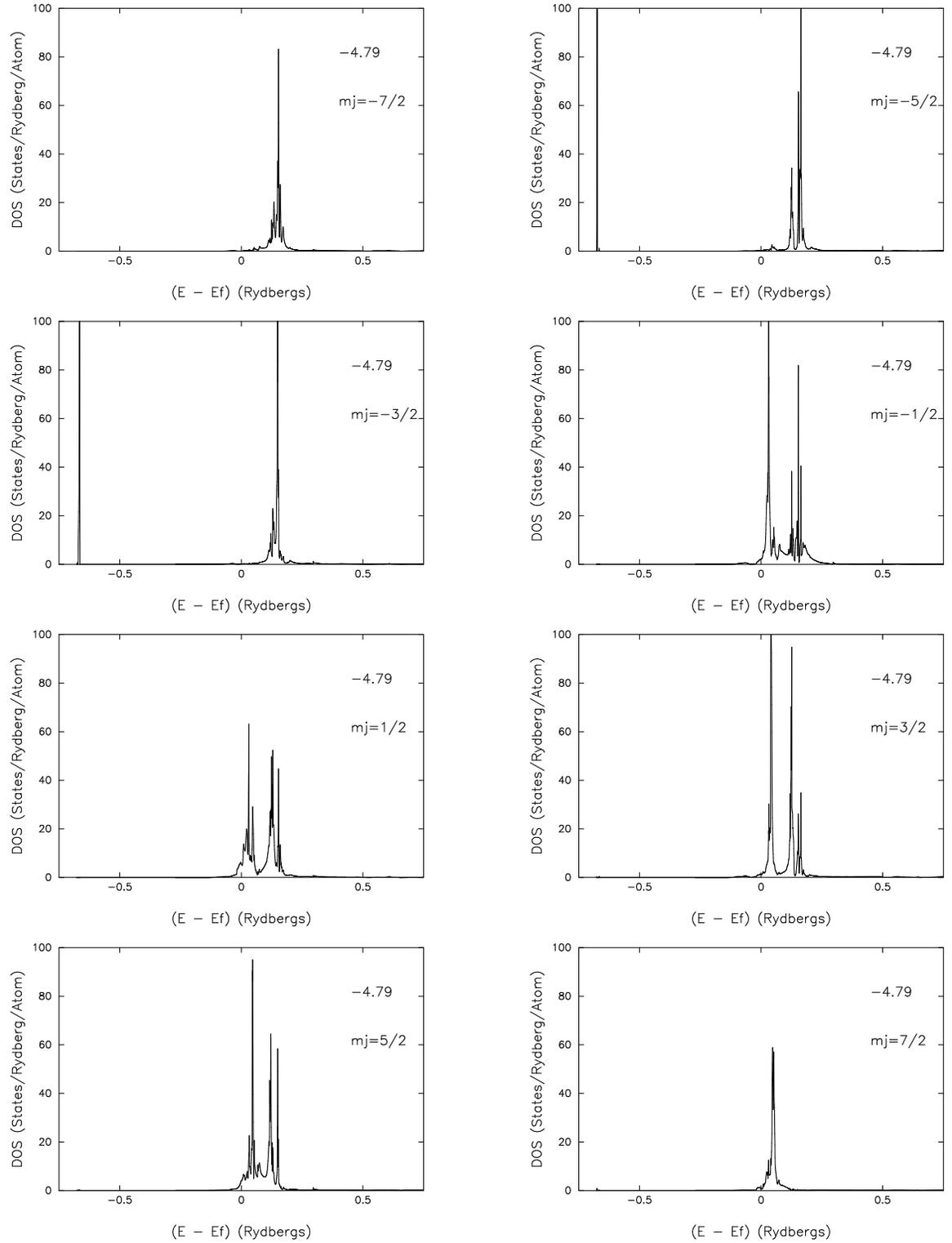

\begin{tabular}{l c c c c c c c c c c c r}
\includegraphics[scale=.33,angle=270]{DOS_DCM5_M7.ps} & & & & & & & & & & & &
\includegraphics[scale=.33,angle=270]{DOS_DCM5_M5.ps} \\
\includegraphics[scale=.33,angle=270]{DOS_DCM5_M3.ps} & & & & & & & & & & & &
\includegraphics[scale=.33,angle=270]{DOS_DCM5_M1.ps} \\
\includegraphics[scale=.33,angle=270]{DOS_DCM5_P1.ps} & & & & & & & & & & & &
\includegraphics[scale=.33,angle=270]{DOS_DCM5_P3.ps} \\
\includegraphics[scale=.33,angle=270]{DOS_DCM5_P5.ps} & & & & & & & & & & & &
\includegraphics[scale=.33,angle=270]{DOS_DCM5_P7.ps} \\
\end{tabular}
\caption{The $l=3$ contribution to the density of states of 
praseodymium decomposed by the $m_j$ quantum number for the case 
when the $m_l=-3$ and $m_l=-2$ $f$ states with spin up are occupied 
(localized). In the top right of each figure is the self-consistently 
calculated orbital moment. Each figure is also labelled with the 
relevant value of the $m_j$ quantum number.}  
\label{fig6}
\end{figure*} 
In relativistic theories of magnetism different values of total 
angular momentum $j$ with the same $z$-component $m_j$ are coupled and 
further decomposition has little meaning \cite{Stra84}. To facilitate 
understanding of the differences in the spectra as orbital moment 
varies we show a selection of density of states curves, decomposed by 
the azimuthal quantum number $m_j$ in Figs. \ref{fig6}--\ref{fig8}. 
There are several points that should be noted about these pictures.

\begin{figure*}
\begin{tabular}{l c c c c c c c c c c c r}
\includegraphics[scale=.33,angle=270]{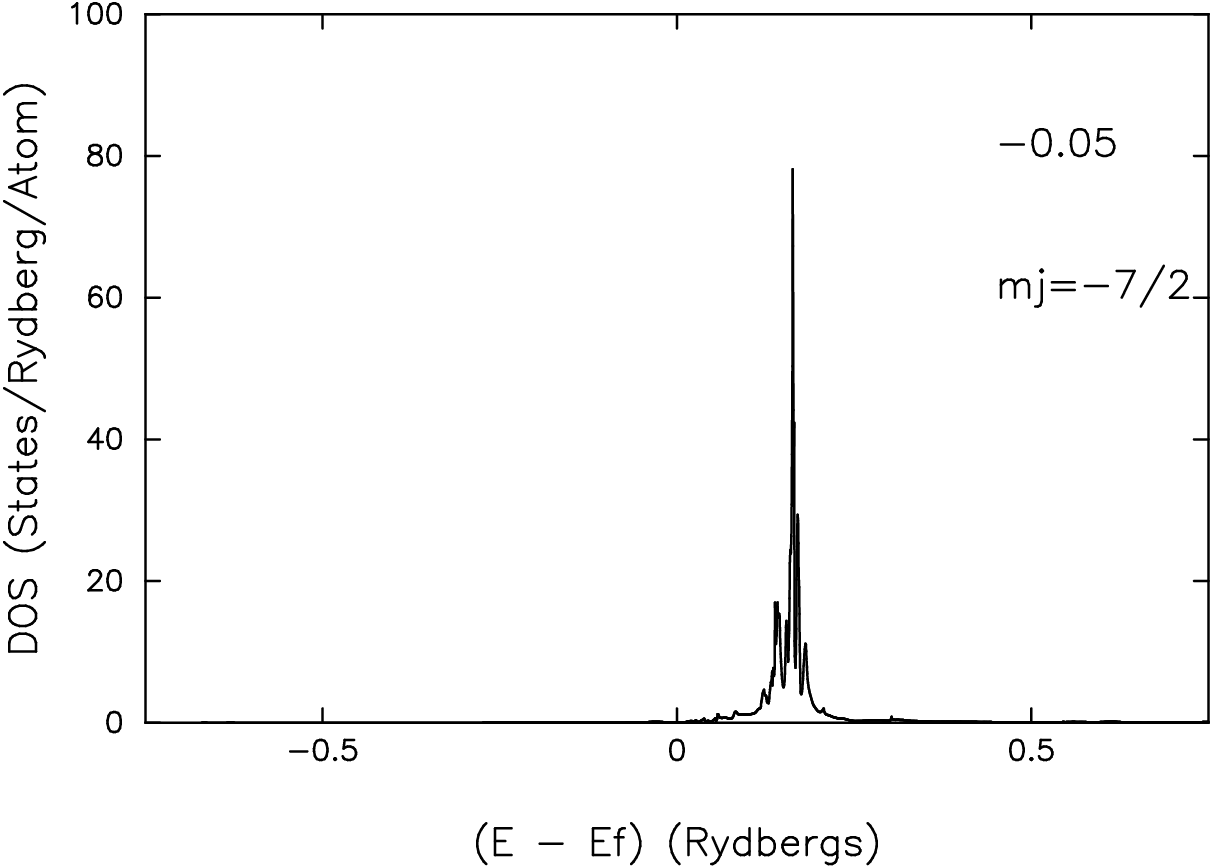} & & & & & & & & & & & &
\includegraphics[scale=.33,angle=270]{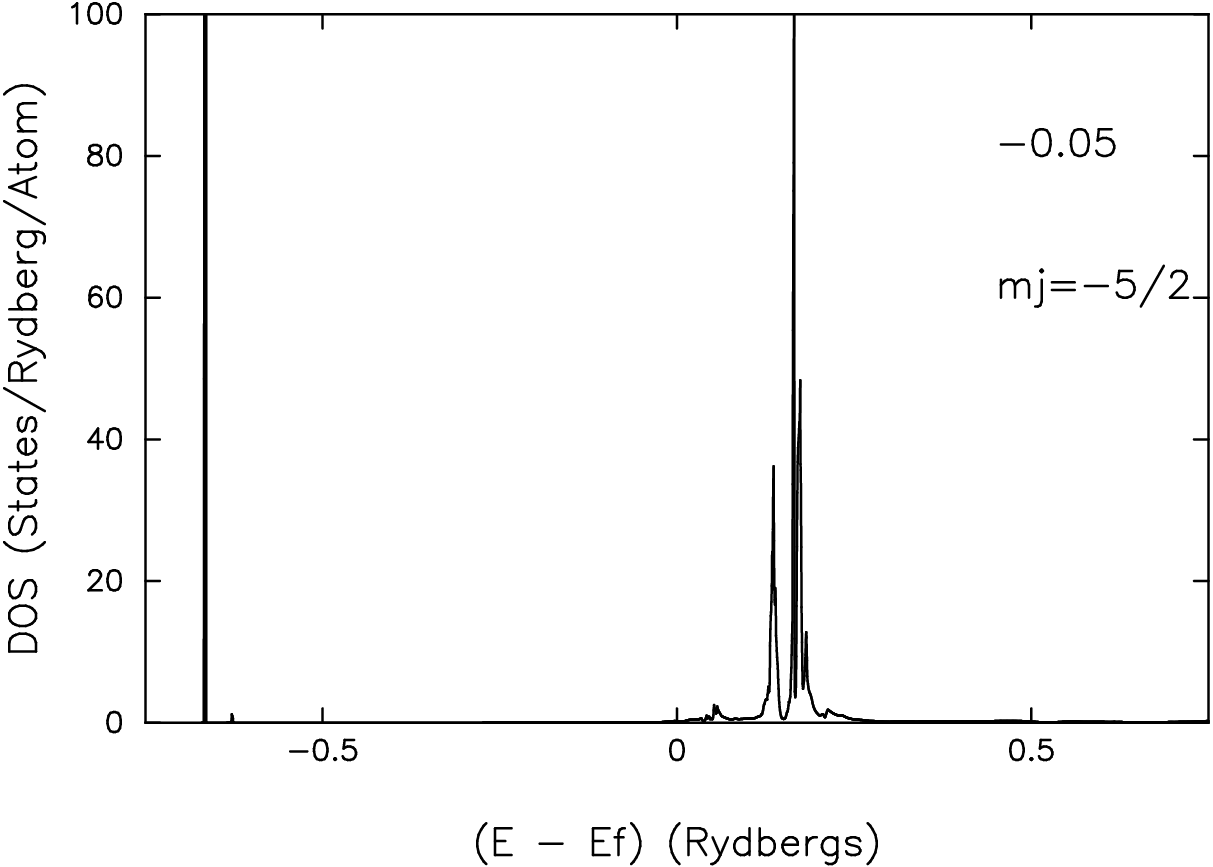} \\
\includegraphics[scale=.33,angle=270]{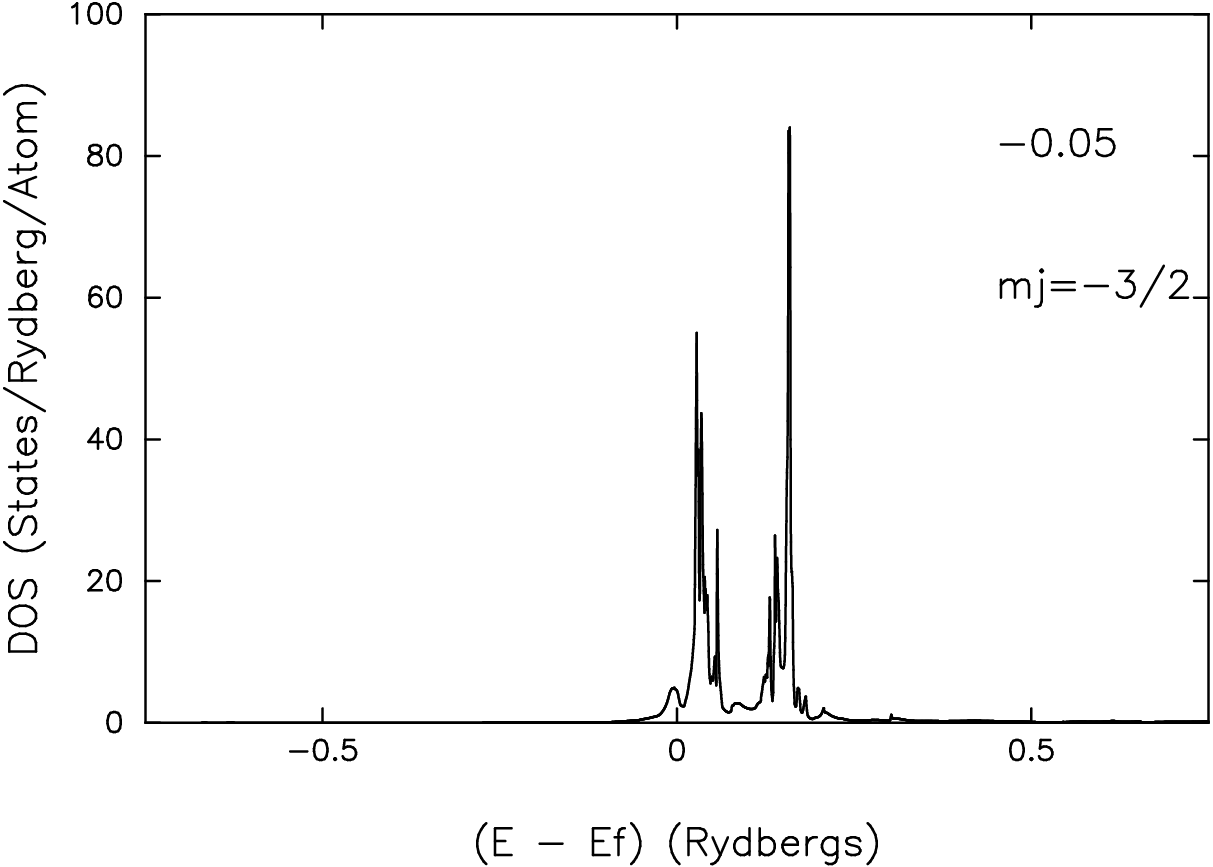} & & & & & & & & & & & &
\includegraphics[scale=.33,angle=270]{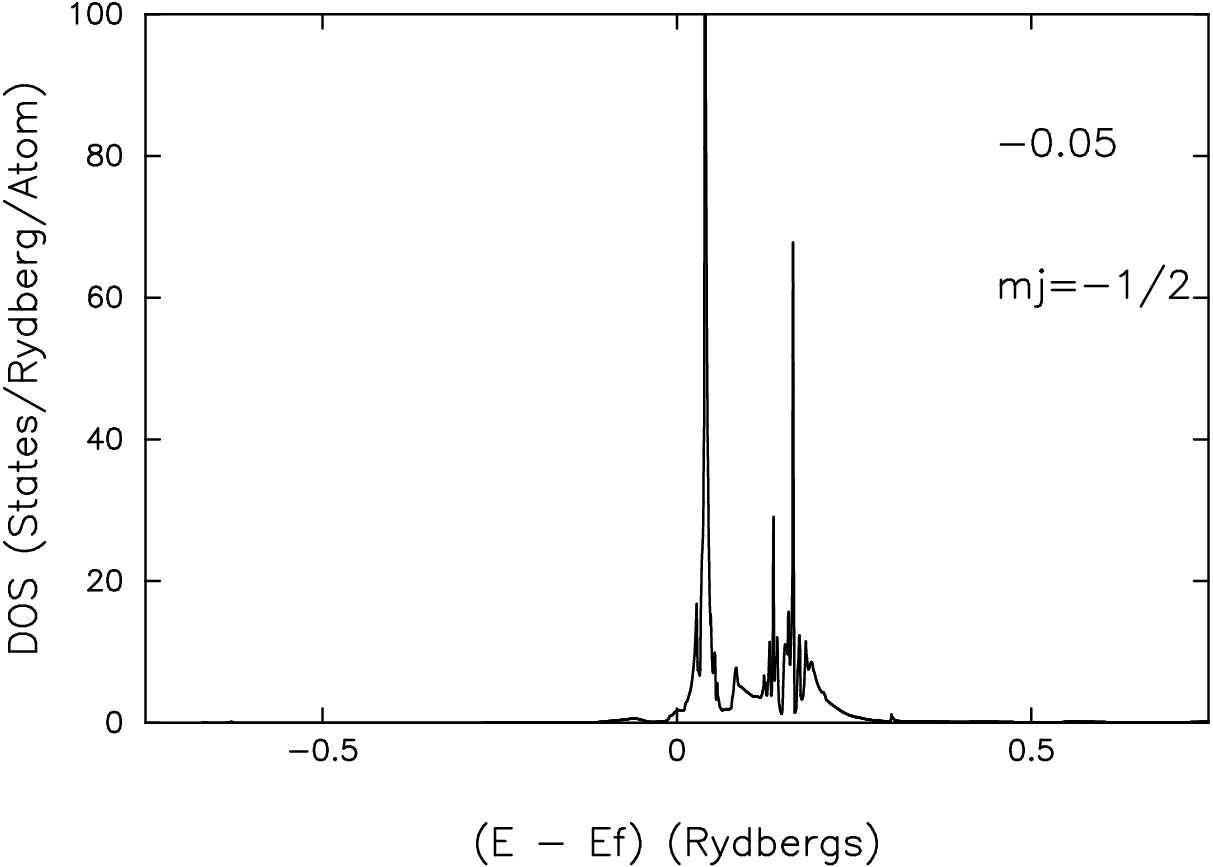} \\
\includegraphics[scale=.33,angle=270]{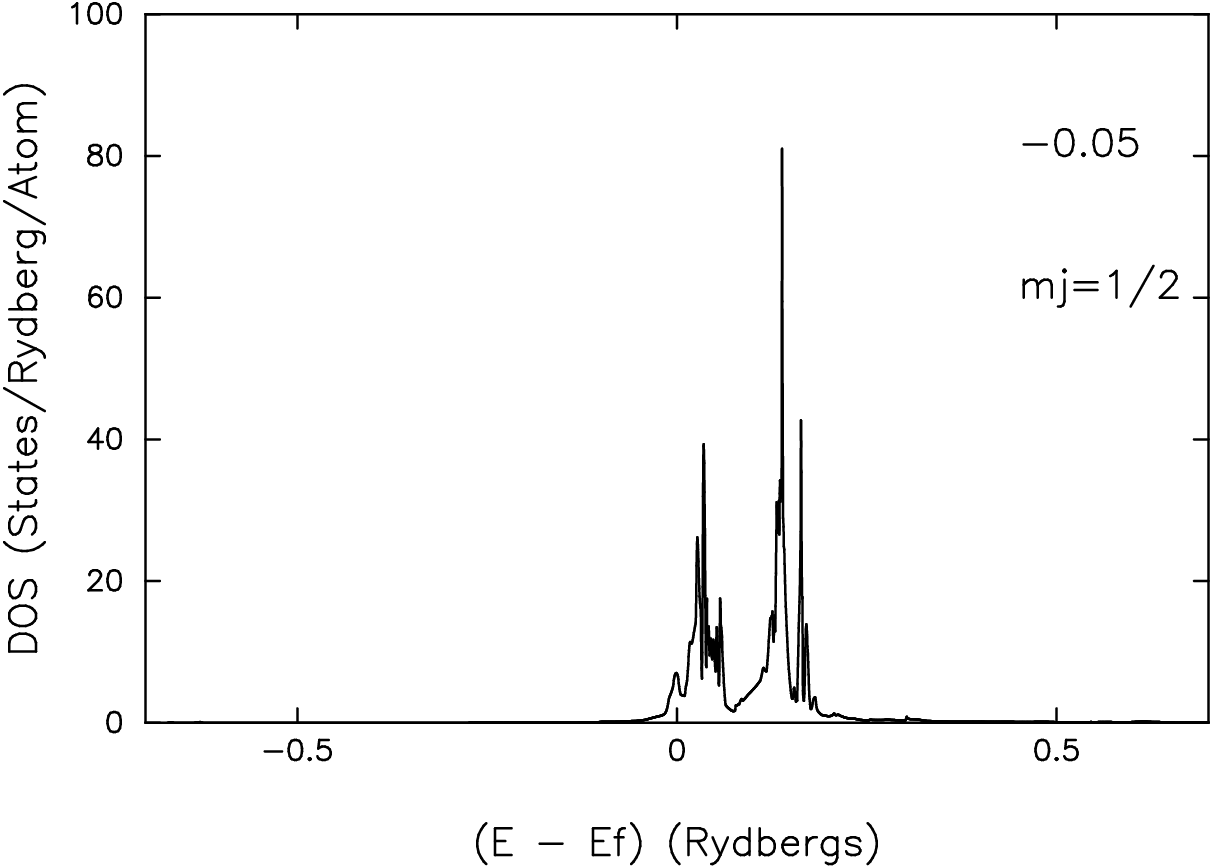} & & & & & & & & & & & &
\includegraphics[scale=.33,angle=270]{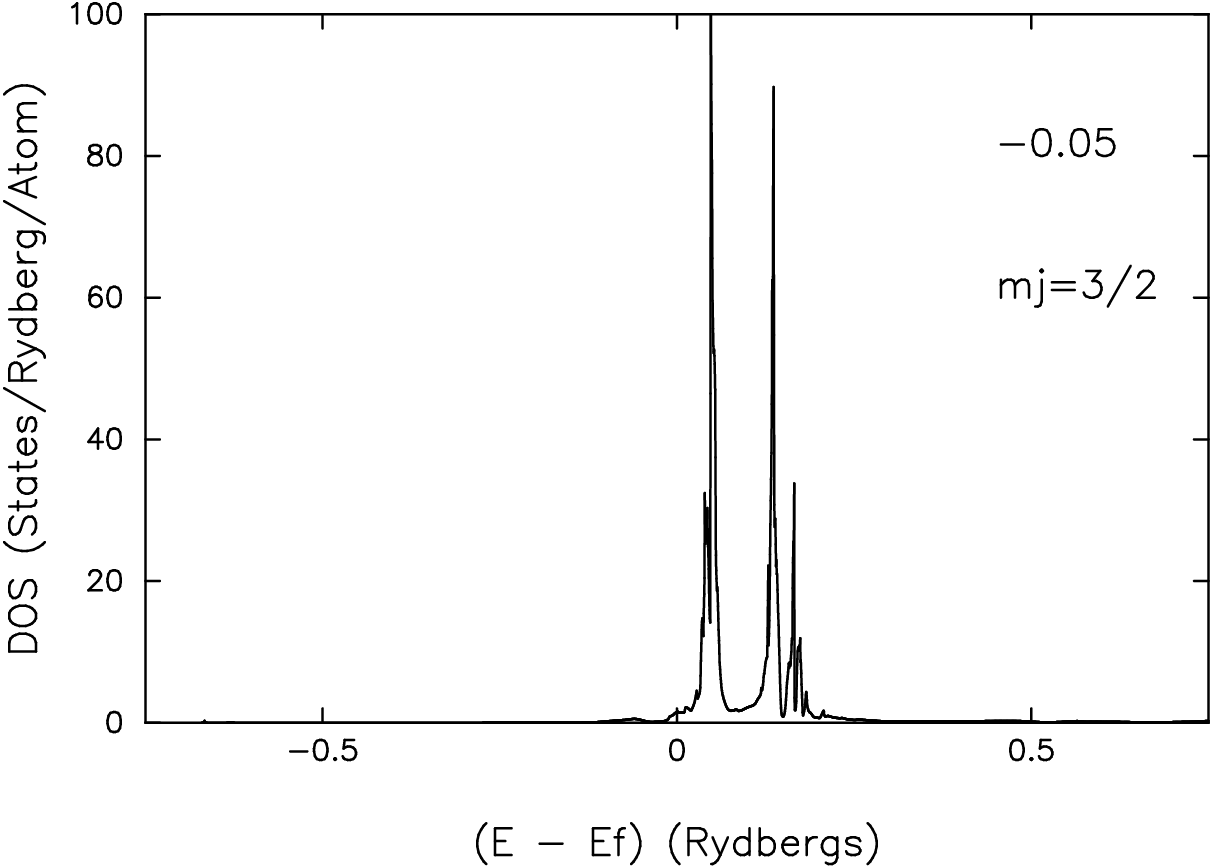} \\
\includegraphics[scale=.33,angle=270]{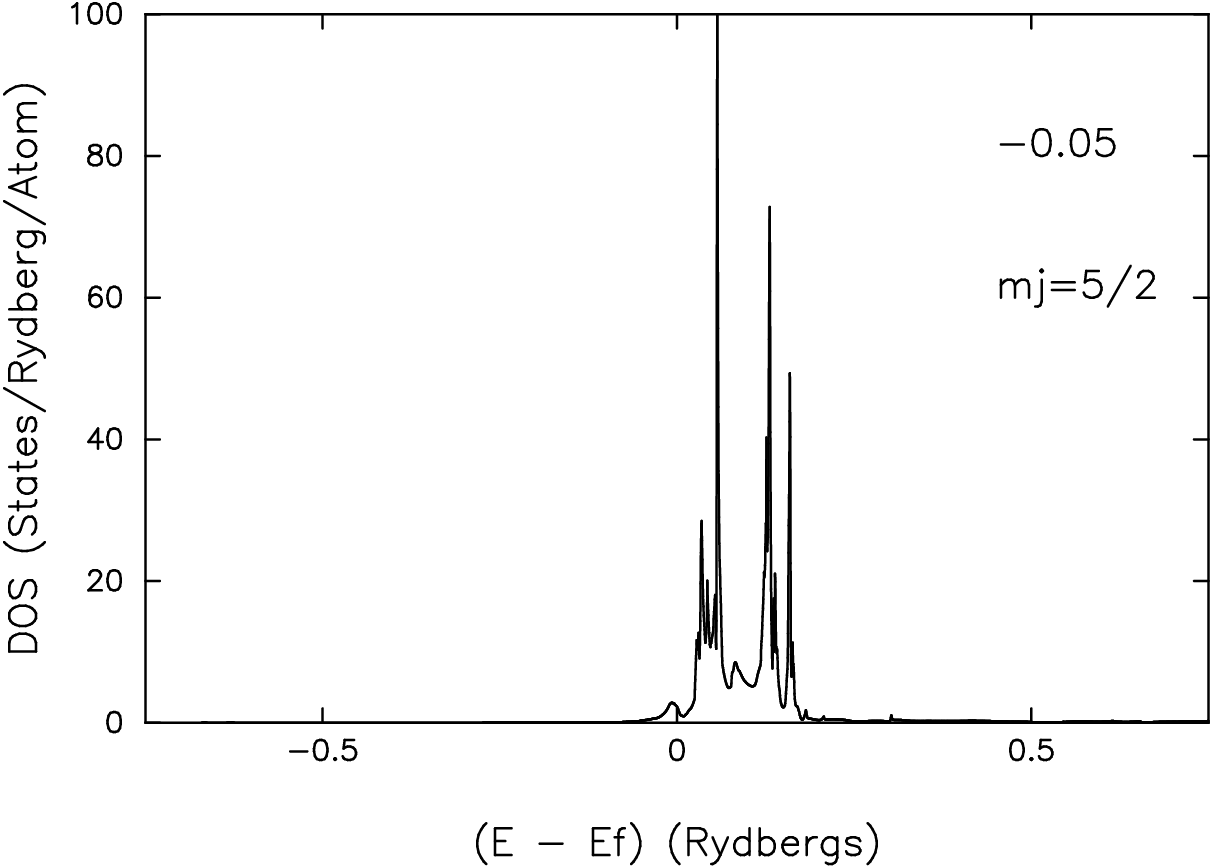} & & & & & & & & & & & &
\includegraphics[scale=.33,angle=270]{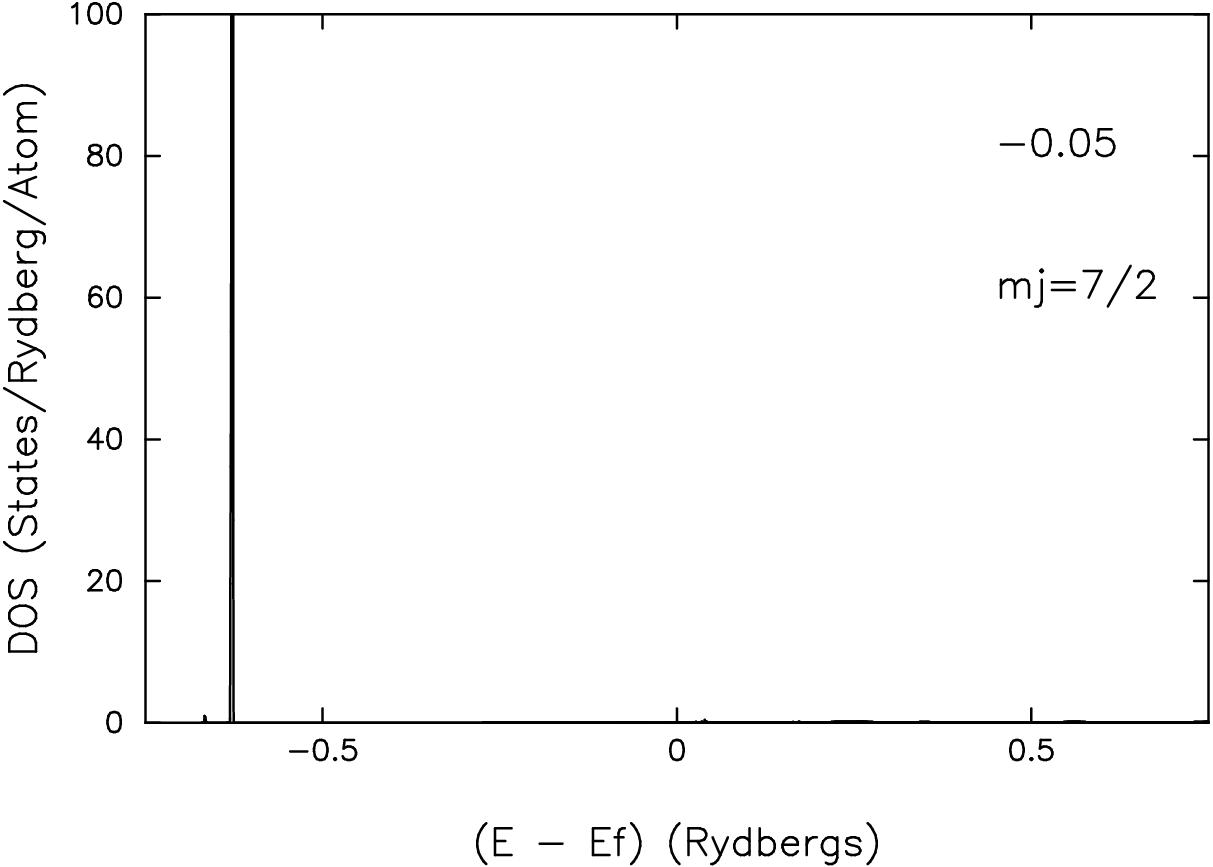} \\
\end{tabular}
\caption{The $l=3$ contribution to the density of states of 
praseodymium decomposed by the $m_j$ quantum number for the case 
when the $m_l=-3$ and $m_l=+3$ $f$ states with spin up are occupied
(localized). In the top right of each figure is the self-consistently 
calculated orbital moment. Each figure is also labelled with the 
relevant value of the $m_j$ quantum number.}  
\label{fig7}
\end{figure*}
The $m_j=\pm 7/2$ (these are pure $j=l+1/2$ states) figures describe 
$f$ electron states with a well-defined $j$ value, while all the 
others show $f$ states with two different values of $j$ ($j=l+1/2$ 
and $j=l-1/2$). In all the pictures except $m_j=\pm 7/2$ there are 
two main peaks, however these two peaks do not necessarily have the 
same weight. The separation of the peaks represents the spin and 
spin-orbit splitting of the individual values of $m_j$. The 
splitting between the unoccupied $f$-states is around 0.1~Ry while 
the splitting between the occupied and unoccupied states is about 
0.7 Ry. The smaller narrow peaks in some of these figures represent 
the hybridization of different $f$-states between themselves. Some 
of these densities of states are markedly broader than others and 
this is a reflection of the degree of hybridization with the 
conduction $s-d$ electrons.

In Fig. \ref{fig6} we have chosen to apply the self-interaction corrections 
to the $f$-electrons which correspond to $(m_l=-3,m_s=+\frac{1}{2})$ 
and $(m_l=-2,m_s= +\frac{1}{2})$ (configuration 1 in Table I) in the 
nonrelativistic limit, and this is reflected in the density of 
states having a very large and narrow peak at around -0.7 Rydbergs 
for $m_j=-5/2$ and $-3/2$. There is nothing for these states to 
hybridize with so they are very tall and narrow atomic-like states. 
For $m_j=-5/2$ and $-3/2$ the density of states has a more 
band-like component corresponding to a single electronic state just 
above the Fermi energy. For most of the other values of $m_j$ there 
is a density of states corresponding to two electron states close to 
$\epsilon_{\rm F}$ and for $m_j=7/2$ the density of states close to $\epsilon_{\rm
F}$ corresponds to a single pure $j=l+1/2$ state. 

In Fig. \ref{fig7} we have selected the $f$-electrons which correspond to 
$(m_l=-3,m_s=+\frac{1}{2})$ and $(m_l=+3,m_s=+\frac{1}{2})$ in the  
nonrelativistic limit for the SIC (configuration 6 in Table I). 
Here it is the $m_j=-5/2$ and the $m_j=+7/2$ components of the 
density of states that have the localized state around -0.7 Rydbergs 
below the Fermi energy. This means there is no $m_j=7/2$ character 
around $\epsilon_{\rm F}$ at all in this case. For most other values of 
$m_j$ we can clearly see that there are two f-states close to 
$\epsilon_{\rm F}$. Detailed examination of these peaks shows that the 
dominant cause of the splitting is the exchange field, although the 
splitting is also influenced by the spin-orbit interaction. For 
$m_j=-7/2$ there is only one state close to $\epsilon_{\rm F}$ of course. 
\begin{figure*}
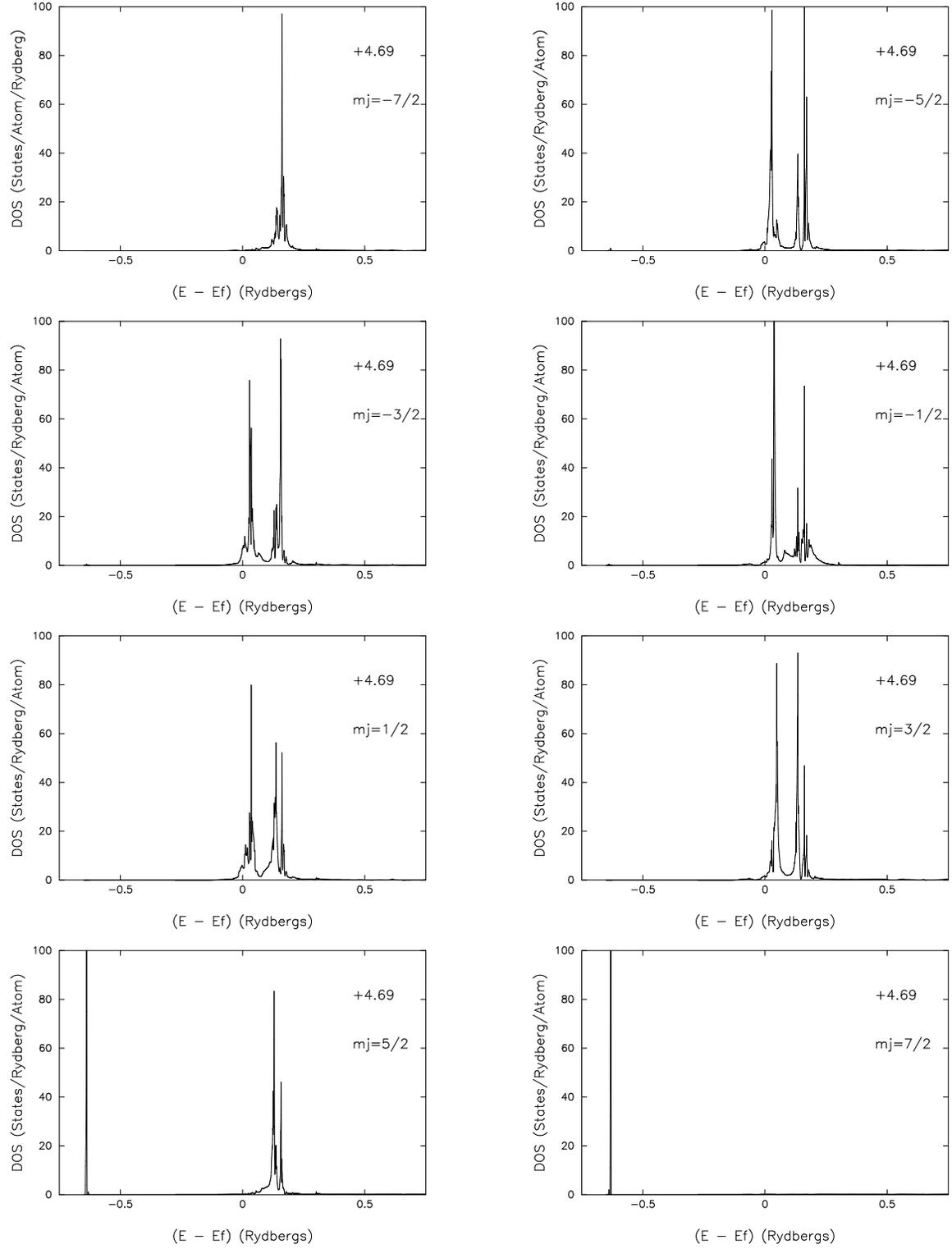

\begin{tabular}{l c c c c c c c c c c c r}
\includegraphics[scale=.33,angle=270]{DOS_DCP5_M7.ps} & & & & & & & & & & & &
\includegraphics[scale=.33,angle=270]{DOS_DCP5_M5.ps} \\
\includegraphics[scale=.33,angle=270]{DOS_DCP5_M3.ps} & & & & & & & & & & & &
\includegraphics[scale=.33,angle=270]{DOS_DCP5_M1.ps} \\
\includegraphics[scale=.33,angle=270]{DOS_DCP5_P1.ps} & & & & & & & & & & & &
\includegraphics[scale=.33,angle=270]{DOS_DCP5_P3.ps} \\
\includegraphics[scale=.33,angle=270]{DOS_DCP5_P5.ps} & & & & & & & & & & & &
\includegraphics[scale=.33,angle=270]{DOS_DCP5_P7.ps} \\
\end{tabular}
\caption{The $l=3$ contribution to the density of states of 
praseodymium decomposed by the $m_j$ quantum number for the case when 
the $m_l=+3$ and $m_l=+2$ $f$ states with spin up are occupied 
(localized). In the top right of each figure is the self-consistently 
calculated orbital moment. Each figure is also labelled with the 
relevant value of the $m_j$ quantum number.}  
\label{fig8}
\end{figure*} 
In Fig. \ref{fig8} we have chosen to apply the self-interaction 
corrections to the $f$-electrons which correspond to 
$(m_l=+3,m_s=+\frac{1}{2})$ and $(m_l=+2, m_s=+\frac{1}{2})$ in the 
nonrelativistic limit (configuration 11 in Table I). This time it is 
the $m_j=5/2$ and $m_j=7/2$ states that are localized, and again there 
is no $m_j=7/2$ character around $\epsilon_{\rm F}$. The $m_j=-5/2$ and 
$m_j=-3/2$ state have the spin-split behaviour close to $\epsilon_{\rm F}$ 
in this case. The other values of $m_j$ behave as before.

It is clear from figures \ref{fig6} to \ref{fig8} that in some $m_j$ 
channels there is a small amount of band-like $f$-character below the 
Fermi energy. This indicates that there are two types of $f$-electron 
in our calculation, the localised $f$-electrons which determine the 
valence and the delocalised $f$-electrons which determine the valence 
transitions. \cite{nature} It is the delocalised $f$-electrons that 
are principally responsible for the non-integer values of the orbital 
moments shown in Figures \ref{fig3} and \ref{fig4}, (although there is 
also a small contribution from the valence $s-d$ electrons).  

Comparison of the corresponding diagrams in Figures \ref{fig6}, 
\ref{fig7}, and \ref{fig8} shows dramatic differences. Even though the 
total density of states is fairly insensitive to which $f$-electron 
states are occupied, the $m_j$-decomposed density of states is obviously 
drastically altered depending on which electrons are localized. In 
particular the $f$-states just above the Fermi energy form a significant 
number of the intermediate states in the formal theory described earlier. 
Therefore if key ones are localized they become unavailable as 
intermediate states for the spectroscopy and the cross section may 
be substantially altered. Of course, occupying one $f$-state means 
that some other $f$ state is not occupied which may then also play 
a role as an intermediate state for the spectroscopy. Indeed, how 
much the unavailability of particular $m_j$ substates affects the 
spectra depends on other factors too, including the $E1$ selection rules 
which are composed of angular matrix elements. Each angular matrix 
element contains four terms in the form of a product of Clebsch-Gordan 
coefficients and a geometry and polarization dependent factor. A 
further influence is the fact that the LMTO coefficients 
$A^{j{\bf k}}_{ti\Lambda}$ (defined in Eq. (17) and completely 
determined by a self-consistent band structure calculation) associated 
with the $f$-electrons are found to be fairly independent of the rare 
earth element under consideration but their magnitude has a clear but 
complex linear proportionality to $m_l$.

Detailed analysis of the major contributions to the cross section 
suggests that the highest peak is formed by the core-to-valence 
transitions  $\left( d_{\frac{3}{2}},m_j \right)\rightarrow 
\left( f_{\frac{5}{2}}, m_j+(-)1\right)$ for the $M_{\text{IV}}$ 
LCP(RCP) edge scattering and $\left(d_{\frac{5}{2}}, 
m_j\right)\rightarrow\left(f_{\frac{7}{2}}, m_j+(-)1\right)$ for 
the $M_{\text{V}}$ LCP(RCP) edge scattering. The former transition 
for $M_{\text{IV}}$ case is in agreement with the nonrelativistic 
selection rule which forbids a ${\Delta}j=2$ transition, although 
this transition is not totally forbidden in the relativistic $E1$ 
selection rule. In the $M_{\text{V}}$ case, the ${\Delta}j=0$ 
transition is observed to form part of the shoulder rather than 
contributing to the main peak. Furthermore, within the transitions 
forming the main peak, the contribution to the LCP scattering at both 
the $M_{\text{IV}}$ and $M_{\text{V}}$ edge is the largest from the
most positive allowed $m_j$ value of the core state. On the other 
hand, the most negative $m_j$ value of the core state gives the 
largest contribution to the RCP scattering. This indicates the fact 
that the Clebsch-Gordan coefficients which are used to calculate the 
selection rules are a dominant factor in determining the relative 
size of the cross section peaks. The origin of this is simply in the 
properties of the Clebsch-Gordan coefficients which vary smoothly between
either 0 and 1 or 0 and -1 depending on the values of the other quantum
numbers.

From these considerations, we see that the separation of the LCP and 
RCP peaks by 1 to 2 electronvolts is a reflection of the 
spin-splitting of the states. In relativistic theory $m_s$ and $m_l$ 
are not good quantum numbers. Furthermore, because of the magnetism, 
different values of $j$ with the same $m_j$ are also coupled. 
However, it is still possible to associate $\langle s_z\rangle$, 
$\langle l_z\rangle$ with these quantum numbers and also to recognize 
the dominant $j$ in atomic-like unhybridized bands. For example, in 
the case of LCP scattering at the $M_{\text{IV}}$ edge, the largest 
contribution to the cross section comes from $(\kappa=3,m_j= 
+\frac{5}{2})$-like orbitals. The two 4$f$ states which have this 
$m_j$ as the main contributor are characterized by $(\langle 
s_z\rangle\simeq +\frac{1}{2}, \langle l_z\rangle\simeq +2)$ and 
$(\langle s_z\rangle\simeq -\frac{1}{2},\langle l_z\rangle 
\simeq +3)$. Electronic structure calculation shows that the former 
state is dominated by $\kappa=-4$ and the latter by $\kappa=3$. 
Therefore the $M_{\text{IV}}$ LCP peak is most affected by the 
availability of the spin-down $\langle l_z\rangle\simeq 3$ state as 
an intermediate state.  Similar analysis shows that the 
$M_{\text{IV}}$ RCP peak is most affected by spin-up 
$\langle l_z\rangle \simeq -3$ state, $M_{\text{V}}$ LCP by spin-up 
$\langle l_z\rangle \simeq 3$, and $M_{\text{V}}$ RCP by spin-down 
$\langle l_z\rangle\simeq -3$ state. 

Although this analysis is a gross simplification, it does explain 
why the relative peak energy positions in the LCP and RCP scattering 
cases swap between the $M_{\text{IV}}$ and $M_{\text{V}}$ edges (see 
Fig. \ref{fig5}). Of course this is true only if these states are still 
available after the chosen localizations by SIC. The effect of 
localization on the MXRS spectrum is most dramatic if SIC is applied 
to these key states, changing the peak energy separation as well as 
the scattering amplitude between the LCP and RCP scattering cases. 

\begin{figure*}
\includegraphics[scale=.70,angle=270]{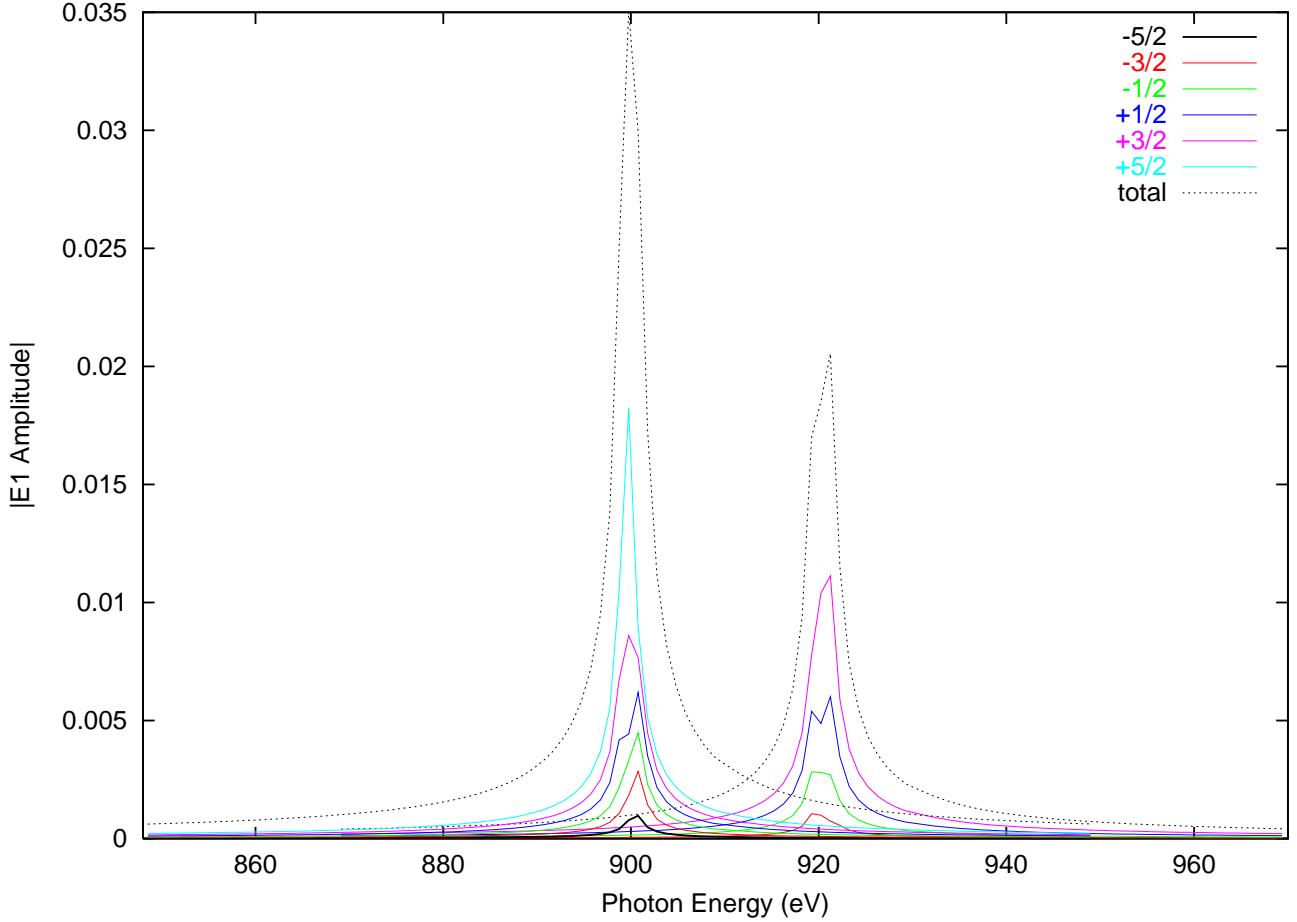}
\caption{Core $m_j$ decomposed LCP amplitude at the $M_{\text{IV}}$ and 
$M_{\text{V}}$ edges.  This figure is for the case when the $m_l=-3$ and 
$m_l=-2$ states are occupied.}
\label{fig9}
\end{figure*}

Some empty valence band $f$ states participating in the scattering 
process have nearly equal mixture of the two $j$ characters, i.e. 
$j=l+1/2$ and $j=l-1/2$. If there is strong spin-up and spin-down 
character in the unoccupied valence states described by a specific 
$m_j$ then both spin states may be available as the intermediate 
states for the spectroscopy. Thus we may clearly see a two-peak 
structure in the $m_j$ decomposed amplitude for a certain 
polarization at the absorption edge. Figure \ref{fig9} shows the core 
$m_j$ decomposed LCP scattering amplitude and the two-peak structure
mentioned above is clearly visible for $m_j=+\frac{1}{2} $ at the 
$M_{\text{IV}}$ edge.

In certain cases we can interpret the apparent relation between the 
magnetic cross section and the $z$-component of the total orbital 
moment as follows.  Because $\langle l_z\rangle+\langle 
s_z\rangle=\langle j_z\rangle$ holds, then we see that if we apply 
self interaction corrections to states systematically according to 
Hund's rules, what is effectively done is to occupy the states in 
order of $m_j$. As stated earlier the $m_j$ decomposed relativistic 
magnetic scattering cross section has a 'proportionality' to $m_j$ 
due to the Clebsch-Gordan coefficient in the angular matrix element 
expression defining the $E1$ selection rules. Whether this 
proportionality is direct or inverse depends on the polarization of 
x rays. In addition, according to the electronic structure calculation, 
as the unhybridized state goes from $(\langle l_z \rangle\simeq -3, 
\langle s_z\rangle\simeq \frac{1}{2})$ to $(\langle l_z\rangle\simeq 
+3, \langle s_z\rangle\simeq \frac{1}{2})$, the dominant $j$ changes 
from $j=\frac{5}{2}$ to $j=\frac{7}{2}$ gradually. This tells us two 
things. Firstly we notice that if a certain state has a major impact 
on the scattering cross section at the $M_{\text{IV}}$ edge for RCP 
photons, then this same state has a relatively minor effect on the 
cross section for LCP photons at the same edge because of the 
Clebsch-Gordan factor in the expression for the $E1$ selection rules 
as mentioned above. Secondly we see that this same state also has 
only a minor effect on the $M_{\text{V}}$ cross section because the 
value of $j$ for the intermediate states involved in major transition 
differ between $M_{\text{IV}}$ and $M_{\text{V}}$. 

As the SIC configuration varies from $(\langle l_z\rangle 
\simeq -3, \langle s_z\rangle \simeq +\frac{1}{2})$ and $(\langle 
l_z\rangle \simeq -2, \langle s_z\rangle \simeq +\frac{1}{2})$ to 
$(\langle l_z\rangle \simeq -3, \langle s_z\rangle\simeq 
+\frac{1}{2})$ and $(\langle l_z\rangle \simeq +3, \langle 
s_z\rangle\simeq +\frac{1}{2})$ so that there is a systematic change 
in the $z$-component of the total orbital moment, the $M_{\text{IV}}$ 
RCP cross section increases because the second, third and so on, 
strongest contributors to the cross section become additionally 
available as intermediate states as they are released from the SIC 
localization. However, they have progressively less impact as we 
proceed through this series of quantum numbers since the major $j$ 
gradually changes to $j=\frac{7}{2}$. The cross section at the 
$M_{\text{IV}}$ edge for LCP photons is not affected much by this 
change in quantum numbers since neither the initial nor the final SIC 
combination in the above series involves the major contributors to 
$M_{\text{IV}}$ LCP cross section. On the other hand, the 
$M_{\text{V}}$ edge LCP cross section is reduced as more and more 
significant contributors are removed from the available intermediate 
states, while the cross section at the $M_{\text{V}}$ edge for RCP 
photons is not much affected for the same reason as $M_{\text{IV}}$ 
LCP case. Obviously the above change in SIC configuration is very 
artificial. However as the states are filled up according to Hund's 
rule as we proceed through the rare earth series, we would expect to 
observe changes in the cross section governed by these considerations 
for rare earths where the intermediate states can be considered as 
atomic-like. However, a very different interpretation of the x-ray 
spectra may be required in the case where delocalized band-like 
intermediate states are of primary importance, as is the case in 
resonant x-ray scattering at the $K$ and $L_{\text{II,III}}$ edges. 

Finally, we are unaware of any experimental measurements of the MXRS 
spectra of praseodymium or it compounds at the $M_{\text{IV}}$ or 
$M_{\text{V}}$ edge. However a careful combined neutron \cite{Goff} 
and x-ray \cite{Vig} (at the $L_{\text{II,III}}$ edges) investigation into 
the magnetism of Ho$_x$Pr$_{1-x}$ alloys has concluded that the Pr 
ion does have a 4$f$ moment at all values of $x$. Deen {\it et 
al.} \cite{Pascal02} have performed MXRS measurements at the $L$ edges 
in Nd/Pr superlattices and found a large peak at the absorption edge 
and a high energy shoulder corresponding to dipolar transitions to 
the broad 5$d$ band. We hope that our calculations will stimulate 
detailed experimental x-ray studies of $M_{\text{IV}}$ and 
$M_{\text{V}}$ edges of Pr, in pure Pr and in its alloys and 
compounds. 
\vskip 5mm

\section{Conclusions}

In conclusion, a theory of magnetic x-ray scattering that is based 
on the LSD with self-interaction corrections and second order 
time-dependent perturbation theory has been described. We have 
illustrated the theory with an application to fcc praseodymium 
and used this example to illustrate the dependence of the scattering 
cross section on spin and orbital magnetic moments. It has been shown that
the theory quantitatively reproduces the dependence on the spin and orbital
magnetic moments originally predicted qualitatively \cite{Blum85}.

\vskip 5mm
\section{acknowledgements}

P. S. and E. A. would like to thank the British EPSRC for research grant 
(number GR/M45399/01) during which the bulk of the work reported in this 
paper was carried out. M. H. would like to thank Keele University for a 
Ph.D studentship. E. A. is also indebted to Tampere City Science 
Foundation for a grant to cover some personal research expenses.


\begin{thebibliography}{}
\bibitem{Eero1} Our definition of the scattering amplitude
[$f^{(\text{our})}(\omega)$] differs slightly from the
standard one [$f^{(\text{std})}(\omega)$] defined by the
differential scattering cross section
$d\sigma/d\Omega=\left|f^{(\text{std})}(\omega)\right|^2$.
They are connected by the relation
$f^{(\text{std})}(\omega)=f^{(\text{our})}(\omega)
V\omega/(2\pi\hbar c^2)$ [see Sec. III A and Ref. 25
in Ref.\ \onlinecite{Eero97}].
\bibitem{Eero3} Single-site solutions of the Dirac
equation with magnetic field
do not have well-defined angular momentum character,
and only $m_j$ remains as a good quantum number
(cf. Eq. (13) of Ref.\ \onlinecite{Eero97}).
\end{thebibliography}
\end{document}